\documentclass[12pt]{article}

\usepackage{graphicx,amsmath,amsfonts,amssymb,amsthm}

\theoremstyle{plain}

\newtheorem{corollary}{Corollary}

\theoremstyle{remark}
\newtheorem{remark}{Remark}
\newtheorem{example}{Example}
\newtheorem{definition}{Definition}

\let\ln=\log

\begin{document}

\title{Thermodynamics as a multistep relaxation process
and the role of observables in different scales of quantities}

\author{V.~P.~Maslov\\
\small Moscow State Institute of Electronics and Mathematics~--\\
\small Higher School of Economics, 109028, Moscow, Russia}

\date{}

\maketitle

\begin{abstract}
In the first part of the paper, we introduce the concept of
observable quantities associated with a macroinstrument measuring the density
and temperature and with a microinstrument determining the radius of a molecule
and its free path length, and also the relationship between these observable
quantities. The concept of the number of degrees of freedom, which relates the
observable quantities listed above, is generalized to the case of low
temperatures. An analogy between the creation and annihilation operators for
pairs (dimers) and the creation and annihilation operators for particles
(molecules) is carried out. A generalization of the concept of a Bose
condensate is introduced for classical molecules as an analog of an ideal
liquid (without attraction). The negative pressure in the liquid is treated as
holes (of exciton type) in the density of the Bose condensate. The phase
transition gas-liquid is calculated for an ideal gas (without attraction).
A comparison with experimental data is carried out.

In the other part of the paper, we introduce the concept of new observable
quantity, namely, of a pair (a dimer), as a result of attraction between the
nearest neighbors. We treat in a new way the concepts of Boyle temperature
$T_B$ (as the temperature above which the dimers disappear) and of the critical
temperature $T_c$ (below which the trimers and clusters are formed). The
equation for the Zeno line is interpreted as the relation describing the
dependence of the temperature on the density at which the dimers disappear. We
calculate the maximal density of the liquid and also the maximal density of the
holes. The law of corresponding states is derived as a result of an observation
by a macrodevice which cannot distinguish between molecules of distinct gases,
and a comparison of theoretical and experimental data is carried out.
In this paper, the observations in three scales, macro, micro, and nano, are studied.
\end{abstract}

\section{Introduction}

When introducing the concept of observable quantity in equilibrium
thermodynamics, one must keep in mind the fact that the observation itself
should be carried out in discrete intervals of time that are widely separated
from one another. When standing on a purely mathematical point of
view\footnote{In due time when the author was constructing
asymptotic expansions of the Schr\"odinger equations in powers of a small
parameter $h$, one of the presently most noted physicists told him
that the asymptotics near the turning points cannot be considered
as semiclassics because the Landau criterion for being semiclassical
is violated there. The author, as a mathematician, believed
that the asymptotics even in the domain of deep shadow
and the ``instantons'' obtained for the imaginary number~$h$
can still be considered as the semiclassical asymptotics.
Recently, Yu.~M.~Kagan clearly explained the author that the physicists
mean only the case $\mu\leq0$ when speaking about the Bose--Einstein distribution.
But the author, as a mathematician, believed that this restriction is artificial
and considered system (1)--(4) in the general case, without any restrictions
on the number $N_i$ of particles at the $i$th energy level.
But the natural restriction $\sum N_i=N$, $N_i\leq N$, still exists
and is taken into account by the author.
The Bose--Einstein condensate also exists
but in a small neighborhood of the zeroth energy level
(small compared with~$n$) rather than at a single point.
The author continues to use the name ``Bose--Einstein''
for the obtained distribution and the condensate phenomenon.
The general asymptotics is constructed for $N\gg\ln N$,
and this asymptotics holds for $N=100$.}, one
must agree that the processes of establishing an equilibrium require infinite
time. However, in mathematics there are some concepts which are similar to the
notion of ``half-life'' in physics. For example, one can introduce a time
interval during which the difference between the current state and the state of
equilibrium in the course of relaxation becomes $e$ times less.

In approximation theory and in the theory of numerical methods, especially
after the well-known paper of Mandel'shtam and Leontovich~\cite{39}, the
following relaxation process was in use: at first, a reacting system is brought
to some equilibrium. Then one rapidly changes one of the conditions (e.g., the
temperature or the pressure) and traces the evolution of the system towards a
new equilibrium (see, for example, the article ``method -- relaxation'' in the
Great Encyclopedia of Oil and Gas, http://www.ngpedia.ru [in Russian]).

Since the observation intervals should be ``equal'' to the relaxation time,
they are large enough, and one can refer to the process as the {\it multi-step
relaxation process} (MRP). Economic and historical processes, and also biological
processes in a living organism, belong to phenomena of this kind, and
therefore, from time to time, thermodynamic models of these processes arise.
The formation of clusters, according to the scheme suggested below in Sec.~4.2, can
serve as an example of a multi-step relaxation process.

The fact that time intervals of observation are discrete is the most important
point to be taken into account when speaking about the instruments of
observation.

The difference between readings of measuring macro- and microinstruments in
thermodynamics is related to the following aspects.

1. A macroinstrument does not take into account the motions of nuclei, of
electrons, and even of atoms within a molecule and regards any molecule as an
individual particle. Mathematically, this corresponds to imposing rigid
constraints on the elements forming the molecule. That is, we must modify those
axioms of mechanics in which we consider all elementary particles and their
behavior in the configuration space whose dimension is equal to the tripled
number of elementary particles.

2. A macroinstrument measuring density counts the number of particles in a
fragment of the volume; however, it cannot trace the movements of particles
with different numbers during discrete finite time intervals. At each discrete
time moment, this device counts the number of particles in the same fragment;
however, it cannot notice what is the exact position of any particle indexed at
the previous time moment and whether or not this particle really is within the
chosen fragment. Mathematically, this means that the arithmetical law of
rearrangement of summands holds. The sum does not depend on the way in which we
have indexed the particles. In this sense, the laws of classical mechanics are
even modified in a more substantial way.

Let us quote from the textbook~\cite{1} on quantum mechanics, where the authors
define the basic property of classical mechanics: ``In classical mechanics,
identical particles (e.g., electrons) do not lose their `personality,' despite
the identity of their physical properties. Specifically, you can imagine that
the particles forming a given physical system are `indexed' at some time moment
and then one can trace the motion of each of the particles along its own
trajectory; then the particles can be identified at any time moment. \dots In
quantum mechanics, it is fundamental that there is no way to trace each of the
identical particles separately and thus to distinguish them. We can say that,
in quantum mechanics, identical particles completely lose their individuality''
(Russian p.~252).

A macroinstrument does not keep this basic property either. Mathematically,
this means that, to take this property into account, we should impose some new
constraints, which are already explicitly nonholonomic, on the mechanics of
many particles and, which is especially important, we should take into account
the permutability of particles in the definition of density, namely, {\it any
permutation of particles does not modify the density}.

In thermodynamics, the gas molecule density is measured.
Although the gas molecules differ from each other
and the Boltzman approach to studying the molecules
is consistent with the objective reality,
the difference between the molecules does not play any role
when the molecule density is determined.
If the density is considered in a small fragment of the vessel,
which contains approximately a million of particles,
then it turns out that the density in this fragment
coincides with the average density in the entire vessel
up to a thousand of particles (up to 0.1 \%)
and is independent of the particle numeration.

It follows from these considerations that the entropy (in contrast to the Boltzmann--Shannon
entropy) should take into account the permutability of the indices of the particles
(cf.~\cite{2}, Sec.~40).

Hence, for an ideal gas
\begin{equation}
\sum_j N_j =\sum_j G_j  \bar{n}_j =N, \qquad
\sum_j \varepsilon_jN_j =\sum_j\varepsilon_j G_j  \bar{n}_j =E,
\tag{1}
\end{equation}
\begin{equation}
\frac{\partial}{\partial\bar{n}_j}(S+\alpha N+\beta E)=0,\tag{2}
\end{equation}
where $\bar{n}_j$ stands for the average number of particles
in each of the $G_i$ states of the $j$th group and
$\alpha$ and $\beta$ are some constants (see~\cite{2}, the footnote on p.~184,
and also~\cite{4} and~\cite{5}),
the entropy must be of the form
\begin{equation}
S=\sum_j  \{(G_j+N_j) \ln (G_j+N_j)-N_j \ln N_j -G_j \ln G_j\},\tag{3}
\end{equation}
\begin{equation}
S=\sum_j G_j [(1+\bar{n}_j) \ln(1+\bar{n}_j) - \bar{n}_j \ln \bar{n}_j].\tag{4}
\end{equation}
In other words, the entropy has exactly the same form as in
the Bose--Einstein quantum case.
This face is proved for balls and boxes in~\cite{2}
in the footnote in Sec.~46; also see~\cite{4,13}.

We have noted above that a macroinstrument and its measurements force
mathematicians to reorganize even the axioms of classical mechanics. However,
mathematicians are forced to do so by entering the corresponding small
parameters and passing to the related limits. A macroinstrument and its
measurements still reduce the time spent to perform constructions of this kind.
However, when one speaks of the axioms of thermodynamics, which is based on
laws derived by great physicists who used ancient experiments conducted on
Earth, it then turns out that the above considerations modify the classical
concept of thermodynamics completely. Meanwhile, microinstruments\footnote{In
mathematics and mechanics, the difference between micro- and macro-observations
is defined as follows: ``the radius $a$ of a molecule is much less than the
typical length of the vessel (provided that the shape of the vessel is
given)'', i.e., there are two scales in the problem, which correspond to macro-
and microinstruments.} also play a role in classical thermodynamics; they
enable one to calculate the dimension related to the number of atoms in the
molecule.

In the mathematical literature, as a rule, the number of degrees of freedom
coincides with the number of independent generalized coordinates. However,
there notions are distinct in the standard thermodynamics, because the volume
is three-dimensional, which is established by the macrodevice, whereas the
number of degrees of freedom is related to the number of atoms in a molecule
and is measured by the microdevice.

Let us explain the following experimental fact. In some
cases, the number of degrees of freedom for diatomic and polyatomic molecules
is an integer.  In our opinion, this happens because the intramolecular communications
(the distances between the atoms of the molecule) are very hard, and, when the
temperature increases, no new degrees of freedom arise. Generally speaking, the
number of degrees of freedom fundamentally depends on the energy of the
molecules, and the energy of different molecules of the same gas is different,
and, apparently, to the average energy (the temperature) there must correspond
the average number of degrees of freedom, which is hence must be noninteger.
However, on one hand, tight connections enable one to excite almost all
molecules for a sufficiently high (room) temperature and, on the other hand, to
give the molecules no possibility to excite new degrees of freedom (e.g., the
vibrational ones). If the connections are not so rigid, then the number of
degrees of freedom depends on temperature, and hence on energy, and is not an
integer in general. This is clear from the comparison of the values of the heat
capacity $C_V$ with the experiment: for hydrogen sulfide with three atoms, the
theory gives 5.96, and the experiment 6.08, for carbon dioxide, the experiment
gives a greater value $C_V=6.75$ ($T=15C$, $P=1 atm $), and, for carbon
disulfide, the vale is almost two times larger, namely, 9.77. In the case of
diatomic molecules, say, for nitrogen, the theory gives 4.967 and the
experiment shows 4.93; for the chlorine, the value is almost 20\% higher,
namely, 5.93, etc.

It turns out that the number of degrees of
freedom coincides with the dimension of the generalized Bose gas which is
regarded as a distribution of a classical gas.

Landau and Lifshitz notice this fact for the three-dimensional Bose gas. They
write that these equations ($PV^{5/3}= \text{const}$) coincide with the
equations of the adiabatic line for an ordinary monatomic gas. ``However, we
stress,'' the authors write further, ``that the exponents in the formulas
$VT^{3/2}=\text{const}$ and $PV^{5/3}=\text{const}$ are not related now to the
ratio of specific heat capacities (since the relations $c_p/c_v=5/3$ and
$c_p-c_v=1$ fail to hold)''~\cite{2}, p.~187.

One can show in a quite similar way that, for the five-dimensional and
six-dimensional Bose gas, the ``Poisson adiabatic line'' coincides with the
Poisson adiabat for the two-atomic and three-atomic molecule (see~\cite{2},
Sec.~47, Diatomic gas with molecules of different atoms. Rotation of molecules).
With regard to the above stipulation, as $\mu\to-\infty$, we obtain precisely
both the condition $c_p-c_v=1$ and the ratio $c_p/c_v$ coinciding with
relations well known in the old thermodynamics.

\begin{remark}
The three-dimensional case of the Bose--Einstein-type distribution
can be represented as
$$
N_j=\sum_{i+k+m=j} N_{i,k,m},
$$
$$
\aligned
M= \sum_{i,k,m} (i+k+m)N_{i,k,m}= \sum_j \sum_{i+k+m=j} (i+k+m)N_{i,k,m}= \\
\sum_j \sum_{i+k+m=j} j N_{i,k,m}= \sum_j j \sum_{i+k+m=j} N_{i,k,m}= \sum_j j N_j.
\endaligned
$$
The Bose--Einstein ``average'' values $\bar{n}_{i,k,m}$
of the occupation numbers $N_{i,k,m}$ depend only on the energy,
i.e., on the sum $i +k+ m$, and
$$
\bar{n}_{i,k,m} =\frac{1}{e^{\beta(i +k+ m-\mu)}-1},
$$
so that
$$
\bar{n}_j =\sum_{i+k+m=j}\bar{n}_{i,k,m}=\frac{q_j}{e^{\beta(i +k+ m-\mu)}-1}, \qquad
q_j=\frac{(j+2)!}{j!3!}.
$$
The transition to integer dimensions is similar;
the fractional dimensions are obtained by passing from factorials
to $\Gamma$-functions. A more rigorous approach in described in~\cite{13,42,43}.
\end{remark}

\section{A new ideal gas and a new ideal liquid\\ as observable quantities}

\subsection{The number of degrees of freedom for $T\leq T_c$ and $P\leq P_c$}

Let us now proceed with finding the number of degrees of freedom for
for a low temperature that does not exceed the critical one:
$T\leq T_c $.

The Maxwell--Boltzmann equation for the ideal gas is of the form
\begin{equation}
PV=NT, \tag{5}
\end{equation}
where $P$ stands for the pressure, $V$ for the volume, $N$ for the number of
particles, and $T$ for the temperature.

Denote by $Z$ the dimensionless quantity $ Z=\frac{PV}{NT},$ which is called
the {\it compressibility factor}. Equation~\thetag{5} can be represented in the
form $Z=1.$ Let us express the Bose--Einstein-type distribution for the fractional
dimension $D$ using polylogarithms.

Represent the thermodynamic potential of the Bose gas of the fractional
dimension $D=2(1+\gamma)$ in the form
\begin{equation}
\Omega(\mu,T)=(Cm)^{1+\gamma}V\frac{T^{2+\gamma}}{\Gamma(2+\gamma)}
\int_0^\infty\frac{t^{1+\gamma}\,dt}{(e^t/a)-1}=
-T^{2+\gamma}(Cm)^{1+\gamma}V\operatorname{Li}_{2+\gamma}(a), \tag{6}
\end{equation}
where $T$ stands for the temperature, $m$ for the mass, $C$ is a constant,
$a=\exp(\mu/T)$ is the activity, $\mu$ is the chemical potential, and $\Gamma$
stands for the Euler gamma function .

The function $\operatorname{Li}_s(a)$ introduced in~\thetag{6} is referred to
as a polylogarithm and is defined by the rule
\begin{equation}
\operatorname{Li}_s(x)=\frac1{\Gamma(s)}\int_0^\infty\frac{t^{s-1}}{(e^t/x)-1},
\qquad \operatorname{Li}_s(1)=\zeta(s), \tag{7}
\end{equation}
where $\zeta(s)$ stands for the Riemann zeta function.

To pass to the dimensionless units, we introduce the temperature $T_r$ in such
a way that $T=T_rT_c$.

The expressions for the dimensionless pressure $P_r=P/P_c$ and for the number
of particles $N$ that correspond to the thermodynamic potential \thetag{6} are
of the form
\begin{equation}
P_r=\frac{T_r^{2+\gamma}\operatorname{Li}_{2+\gamma}(a)}{\zeta(2+\gamma)},\tag{8}
\end{equation}
\begin{equation}
N=V T_r^{1+\gamma}\operatorname{Li}_{1+\gamma}(a).\tag{9}
\end{equation}
We have (for the definition of~$\gamma_c$, see below)
\begin{equation}
\Omega'=-T_r^{2+\gamma}(Cm)^{\gamma_c-\gamma}V\operatorname{Li}_{2+\gamma}(a).
\tag{10}
\end{equation}

The following formula can thus be obtained for the compressibility factor $Z$:
\begin{equation}
Z=\frac{\operatorname{Li}_{2+\gamma}(a)}{\operatorname{Li}_{1+\gamma}(a)}.
\tag{11}
\end{equation}

In particular, for $a=1$ (i.e., for $\mu=0$), we have
\begin{equation}
Z=\frac{\zeta(\gamma+2)}{\zeta(\gamma+1)}. \tag{12}
\end{equation}

As is well known, in the Bose--Einstein theory, the value $\mu=0$ corresponds
to the so-called degeneration of the Bose gas.

For a classical gas satisfying the same relations, the {\it degeneration
coincides with the critical point\/} $T=T_c$, $P=P_c$, and $Z=Z_c$.
Consequently, one can write $\gamma=\gamma_c$ for $Z=Z_c$ in~\thetag{12},
namely,
\begin{equation}
Z_c= \frac{\zeta(\gamma_c+2)}{\zeta(\gamma_c+1)}, \tag{13}
\end{equation}
and to every pure classical gas there corresponds its own value of $\gamma_c$.

The entropy in the dimension $D=2\gamma +2$ can be evaluated in the standard
way. The great thermodynamical potential is considered,
$$
\Omega= -PV =-\frac{VT}{\Lambda^{2(1+\gamma)}}\cdot \frac{1}{\Gamma(2+\gamma)}
\int_0^{\infty}\frac{t^{1+\gamma}\, dt}{(e^t/a)-1}
=\frac{-VT^{2+\gamma}}{{\Lambda'}^{2(1+\gamma)}}\operatorname{Li}_{2+\gamma}(a),
$$
where $\Lambda' = \text{const}/(2\pi m)^{1/2}$, the dimension~$D$ is equal to
$2\gamma +2$, $T$ stands for the temperature, and $a= e^{\mu/T}$  for the
activity.

The number of particles is
$$
N=-\frac{\partial\Omega}{\partial\mu}=\frac{V T^{1+\gamma}}{{\Lambda'}^{2(1+\gamma)}}
\operatorname{Li}_{1+\gamma}(a).
$$
The compressibility factor is
$$
Z=\frac{PV}{NT}=\frac{\operatorname{Li}_{2+\gamma}(a)}
{\operatorname{Li}_{1+\gamma}(a)}.
$$
Let us evaluate the entropy,
\begin{align*}
S=-\left(\frac{\partial\Omega}{\partial T}\right)_{V,\mu}
&= (2+\gamma) \frac{V
T^{1+\gamma}}{{\Lambda'}^{2(1+\gamma)}}\operatorname{Li}_{2+\gamma}(a) -
\frac{V T^{1+\gamma}}
{{\Lambda'}^{2(1+\gamma)}}\operatorname{Li}_{1+\gamma}(a)\frac{\mu}{T}
\\
&= \frac{V T^{1+\gamma}}{{\Lambda'}^{2(1+\gamma)}}
\left[(2+\gamma)\operatorname{Li}_{1+\gamma}(a)
-\operatorname{Li}_{1+\gamma}(a)\frac{\mu}{T}\right].
\end{align*}
For $\mu=0$, $T_r=T/T_c$, and $P_r=P/P_c$, the specific entropy is equal to
\begin{equation}
\frac SV\big|_{\mu=0, T_r=1}=
(2+\gamma)\zeta(2+\gamma).\tag{14}
\end{equation}

E.~M.~Apfel'baum [Apfelbaum] and V.~S.~Vorob'ev~\cite{3} compared the Bose
distributions of fractional dimension $D=2\gamma_c+2$ in the $(P,V)$ diagram
with the experimental critical isotherms for various gases. We present these
graphs in Figs.~1--5. In Fig.~5A, the graphs for the nitrogen and oxygen are
shown, which have been constructed by Professor V.~S.~Vorob'ev.

\begin{figure}[h!]
\begin{center}
\includegraphics{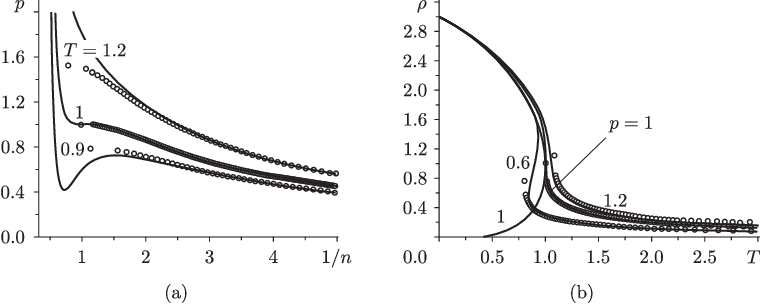}
\end{center}
\caption{(a) Isotherms of pressure for the van der Waals equation are shown by solid
lines. The lines formed by circles are constructed from computations for
$\gamma=0.312$ (i.e., for the ideal ``Bose gas''), $Z_{\mathrm{cr}}=3/8$,
$p=P/P_{\mathrm c}$, and $n=N/N_{\mathrm c}$. \newline (b) Isobars of density
for the van der Waals equation are shown by solid lines. Line 1 is the binodal.
The circles correspond to isobars of the ``Bose gas'' for $\gamma=0.312$.}
\end{figure}

\begin{figure}[h!]
\begin{center}
\includegraphics{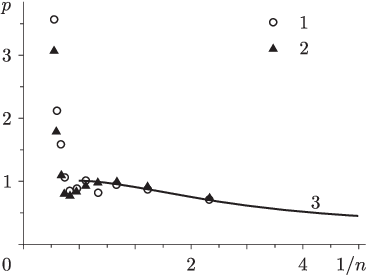}
\end{center}
\caption{Critical isotherms for the Lennard--Jones system.
Symbols~1 and~2 correspond to numerical calculations.
Line~3 corresponds to the ideal Bose gas for
$\gamma=0.24$.}
\end{figure}

\begin{figure}[h!]
\begin{center}
\includegraphics{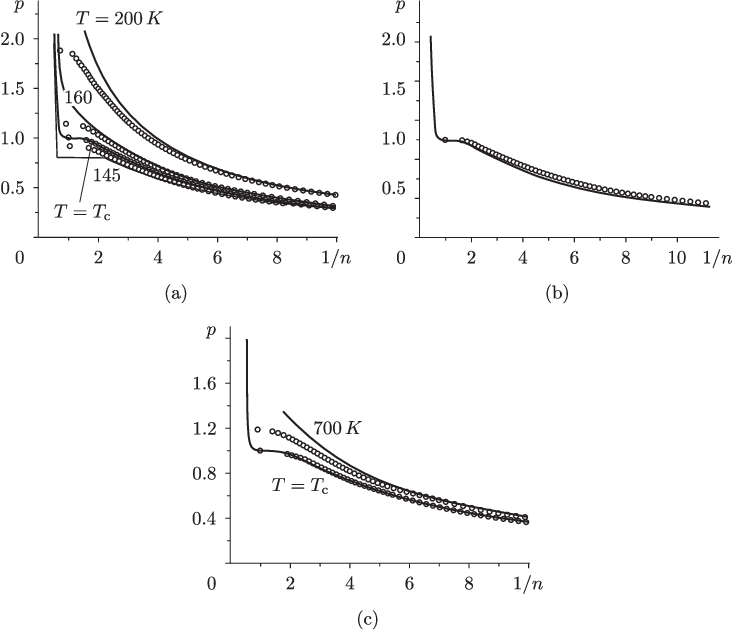}
\end{center}
\caption{(a) Isotherms for argon. The solid lines correspond to experimental data.
The line formed by circles is constructed in accordance with the isotherm of the
ideal ``Bose gas'';
$Z_{\mathrm{cr}}=\frac{\zeta(\gamma+2)}{\zeta(\gamma+1)}=0.29$, $p=P/P_{\mathrm
c}$, and $n=N/N_{\mathrm c}$. \newline (b) The same for water,
$Z_{\mathrm{cr}}=0.23$. \newline (c) The same for copper,
$Z_{\mathrm{cr}}=0.39$. }
\end{figure}

\begin{figure}[h!]
\begin{center}
\includegraphics{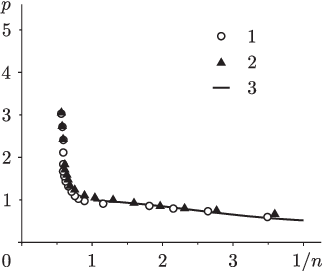}
\end{center}
\caption{Isotherms for water. Symbols 1 and 2 correspond to experimental data, and
line~3 corresponds to the computation for the Bose gas.}
\end{figure}

\begin{figure}[h!]
\begin{center}
\includegraphics{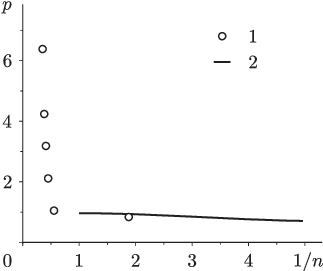}
\end{center}
\caption{Critical isotherms for mercury. Symbols~1 and~2 correspond to experimental
data, and line~3 corresponds to the computation for the Bose gas.}
\end{figure}

\setcounter{figure}{4}
\renewcommand{\thefigure}{\arabic{figure}A}

\begin{figure}[h!]
\begin{center}
\includegraphics[width=7.5cm]{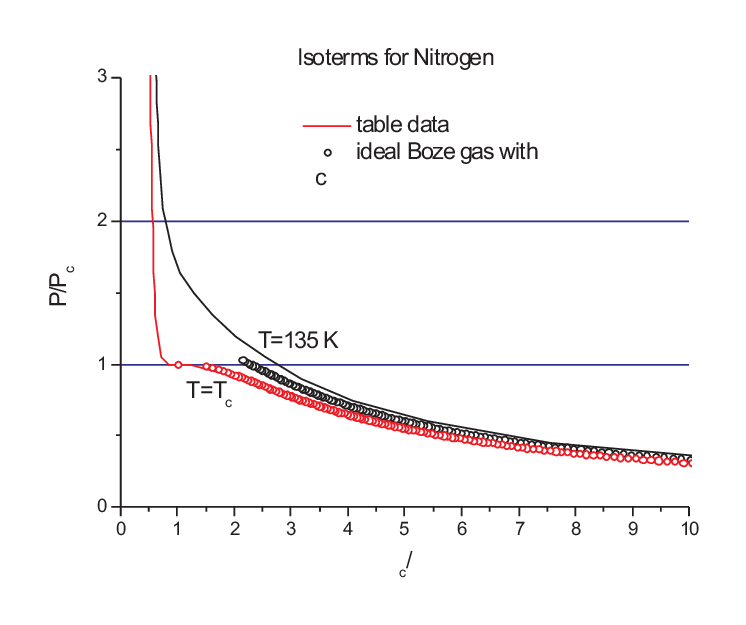}
\includegraphics[width=7.5cm]{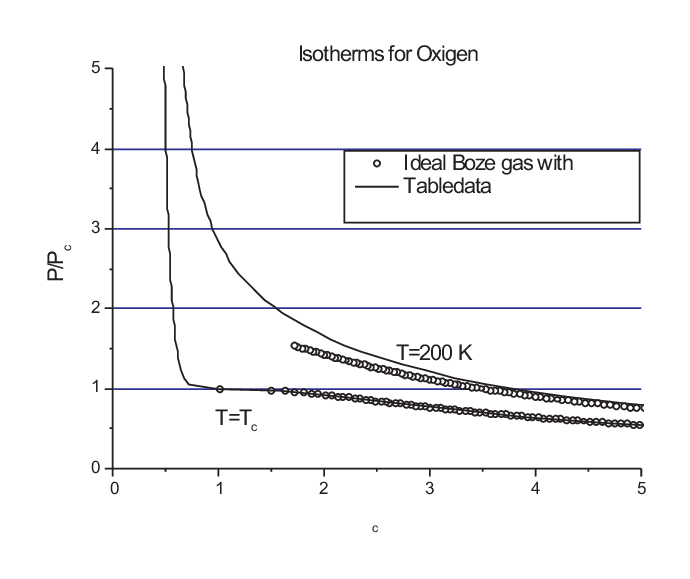}
\end{center}
\caption{1. Isotherms for nitrogen. 2. Isotherms for oxygen.}
\end{figure}

\newpage
\clearpage

\subsection{Bose condensate as an observable quantity\\ in classical thermodynamics.
Relativity principle for MRP}

We shall show that the Bose condensate in classical thermodynamics is the
condensate of gas (vapor) into liquid (in contrast to the statement presented
in the manual~\cite{2} in the footnote on p.~199).
\medskip

{\bf Example 1.} Consider the example given by a famous theorem in number
theory, namely, the solution of an ancient problem, which has the Latin title
``partitio numerorum.'' This task involves an integer $M$ which is decomposed
into $N$ terms, for example, if $M=5$ and $N=2$, then
$ 5 = 1 +4 = 2 +3,$ which gives two solutions to the problem, $\mathcal{M} = 2$.

If $M=10^{23}$ and $N = 1$, then the decomposition has only one version, and
$\mathcal M = 1$. If $M=10^{23}$ and $N=10^{23}$, then there is also only one
version of decomposition, namely, the sum of ones, i.e., $\mathcal{M} = 1$.

Obviously, there is a number $N_c$ for a fixed $M$ such that the number of
versions of the decomposition, $\mathcal M$, is maximal possible (this number
is not unique in general). The number $\log_2\mathcal M$ is referred to as the
{\it Hartley entropy}. At the point at which $\mathcal{M}$ reaches its maximum,
there is a maximum entropy.

Let a partition $ M=a_1+\dotsb+a_N$
of $M$ into $N$ summands be given. Denote by $N_j$ the number of summands on
the right-hand side that are precisely equal to~$j$.

Then the total number of summands is $\sum_jN_j$, and this number is equal to
$N$, since we know that the total number of summands is $N$. Further, the sum
of the parts equal to $j$ is $jN_j$, since there are $N_j$ summands, and then
the sum of all summands can be obtained by summing these expressions over all
possible $j$, i.e., $\sum_j jN_j$, and this sum is equal to $M$. Namely,
\begin{equation}
\sum_{i=1}^\infty N_i=N, \qquad \sum_{i=1}^\infty iN_i=M. \tag{15}
\end{equation}

The very nonuniqueness of the above maximum and an uncertainty concerning the
number of the maxima enabled Erd\H{o}s to obtain results with accuracy up to
$o(\sqrt M)$ only.

Thus, the Erd\H{o}s theorem holds for the system of two Diophantine equations
\begin{equation}
\sum_{i=1}^\infty N_i=N, \qquad \sum_{i=1}^\infty iN_i=M. \tag{16}
\end{equation}
The maximum number of solutions of the system is achieved provided that the
following relation holds:
\begin{equation}
N_c=\beta^{-1}M_c^{1/2}\log\,M_c+\alpha M_c^{1/2}+o(M_c^{1/2}), \qquad
\beta=\pi\sqrt{2/3}, \tag{17}
\end{equation}
and the coefficient $\alpha $ is defined by the formula
$\beta/2=e^{-\alpha\beta/2}$.

If one increases the number~$N$ in problem~\thetag{16} and keeps the number~$M$
constant, then the number of solutions decreases. If the sums in~\thetag{16}
are counted from zero rather than from one, i.e.,
\begin{equation}
\sum_{i=0}^\infty iN_i=(M-N), \qquad \sum_{i=0}^\infty{N_i}=N, \tag{18}
\end{equation}
then the number of solutions does not decrease and remains constant.

Let us explain this fact. The Erd\H{o}s--Lehner problem~\cite{6} is to decompose
a number $M_c$ into $N\leq N_c$ summands.

The decomposition of the number 5 into two summands has two versions. If we
include also 0, then we obtain three versions, 5+0 = 3+2 = 4+1. Thus, the
inclusion of zero gives the opportunity to say that we decompose the number
into $k\leq n$ summands. Indeed, the expansion of the number 5 into 3 summands
includes all previous versions: 5 +0 +0, 3 +2 +0. and 4+1 +0 and adds new
options that do not contain zero.

Here the maximum does not change much~\cite{6}; however, the number of options
cannot decrease, because the zeros enable the maximum to remain constant, and
the entropy never decreases; after reaching the maximum, it becomes constant.
This very remarkable property of entropy enables us to construct a general
unbounded probability theory~\cite{7}. In physics, the effect is identical to
the so-called phenomenon of Bose condensate.

Let us pose the following question: what is the difference between arithmetic,
together with the problem of ``partitio numerorum,'' and the Boltzmann--Shannon
statistics? If we assume that $4+1$ and $1+4$ are two different versions, then
we obtain the Boltzmann--Shannon statistics. The number of versions of
decomposition, $\mathcal M$, is growing rapidly. Thus, the ``noncommutativity''
of the addition gives additionally a huge number of versions of decomposition,
and the Hartley entropy (which is equal to the logarithm of the number of
versions) coincides with the Boltzmann--Shannon entropy.
\endexample

Therefore, we have proved that, if we add zero to the family of possible
summands and decompose a number $M$ into $N$ summands, then this is equivalent
to solving equations~\thetag{18},
i.e., to imposing relations for the number of particles and for energy. Here
the number of zeros increases drastically; if $M=5$, then, for $N>3$, the
number of zeros is 22.
However, the number of ones is also large,
although it is twice smaller than the number of zeros.
\medskip

It is very visible to consider the Bose condensate as the number of zeros;
however, this is inaccurate. The Bose condensate occurs in a neighborhood of a
point at which the energy vanishes rather than at the point itself.
Nevertheless, if one writes $\rho_0\delta(\mathbf k)$ (where $ \rho_0 $ stands
for the density and the vector $\mathbf k$ is the momentum) for the Bose
condensate at rest, then this notation is true, because the density is the
limit
$$
\rho_0=\lim_{N\to\infty,V\to\infty}\frac{Nm}V,
$$
where $N$ stands for the number of particles, $V$ for the volume, and $m$ for
the mass of the particle. This means that, as $ N\to\infty $, the bell-shaped
function near the zero energy is converted to the $\delta$ function.

By an ideal (or perfect) liquid we mean a liquid without attraction and without
any surface tension. This is a liquid which can exist for a positive pressure
only together with a saturated steam.

In this case, the perfect liquid is the result of an optical illusion, and this
``perfect liquid'' is the same ideal gas in the condensate, with another
density. It can be described, as in the case of a consideration of a Bose
condensate, in the form $\rho_0\delta(\mathbf k)$, where $\rho_0$ stands for the
density of the condensate. It cannot exist without a volume trap, which is
similar to the case in which a container with gas has a hole, and there is a
vacuum outside the vessel, in which case this liquid, which looks as if it is
boiled, is going away together with the gas. The mean speed of the particles
inside the liquid is the same as the mean speed of the gas particles. This
corresponds to the condition that the temperature in the system
``liquid--saturated steam'' is the same in the liquid and in the gas. The
liquid in a closed vessel is a fluctuation standing at a fixed place
($\delta(\mathbf k)$), or, speaking in a simpler way, this liquid is a
``resting'' Bose condensate (cf.~\cite{8}, p.204).

Small crystals that occur in a supersaturated solution coincide with the Bose
condensate only if they are not composed of mutually connected particles, and,
moreover, if the particles are continuously exchanged with the particles in the
solution; moreover, the small crystals, as solids, are an optical illusion,
namely, we simply do not see that the particles of crystals are permanently
transposed with particles of the solution. In other words, this is by no means
a crystal, this is a fluctuation; however, this fluctuation is relatively
immobile.

Thus, the Bose condensate for classical particles represents some ``special
density fluctuations;'' only this, and nothing more.

A.~I.~Anselm constructed his theory starting from the Eyring formula
for free energy. The liquid structure model accepted by Eyring is in fact
closer to the strongly compresses gas model~\cite{44,45}.

One can talk about the
density in a ``special cluster fluctuation'' of a part of our vessel with gas.
If we speak of the density in this cluster only, this means that (as in the
example of a small volume with one million of particles) one cannot speak of
the number of particles that are placed in the cluster as if they are frozen
and do not move. This is only an appearance, and all of the particles or a part
of them  can be replaced in a minute by another ones, and the indexing inside
the ``fluctuation'' cluster can change every minute.
At the next time step, this can be the same picture but probably
with different particles involved.

We speak about some fragment of the volume. In fact, the particles that are
more concentrated can be spread out over the entire vessel. However, if there
is at least a little gravity of the Earth, then the fluctuations with more
concentrated particles accumulate near the bottom. If we consider a vessel with
gas in the form of a perfectly reflecting sphere (see~\cite{9}--\cite{12}), then, due
to the repulsive force occurring at the border, fluctuations of this kind are
located near the center of the ball.

From the standpoint of our observation,
in discrete time intervals at far distances from each other,
the denser fragment of the volume, i.e., the Bose condensate,
is at rest and hence corresponds to a small momentum
in the Bose--Einstein-type distribution.
Mathematically, the MRP model corresponds to this phenomenon.
This property will be called the \textit{relativity principle for MRP}.

Let us repeat once again that the only fact which can be guaranteed by the
generalized theory of Bose condensate is that there will be a higher density
of particles at the bottom.
\medskip

{\bf Example 2.} Let a gas be contained in a closed vessel at a room
temperature, and let the gas be almost satisfying the Clausius relation
\begin{equation}
P=\rho T. \tag{19}
\end{equation}
We cool the vessel down to a temperature $ T = 0 $. At some temperature $T\geq
T_0$, a liquid is formed. The temperature $ T_0 $ is referred to as the dew
point. According to the standard conception, the fluctuations above the
temperature of the dew point are of the order of $\sqrt{N}$. After the
formation of liquid, the gas, which is called a saturated steam in the physical
literature, also satisfies relation~\thetag{19}. It is quite rarefied.
According to the van der Waals model, there are no singularities at the dew
point under the gas-liquid passage (on the so-called binodal). According to
experimental data, there are no large fluctuations either in the usual sense at
the dew point.
\medskip

Finally, the most important thing. The experiment shows that, at $ T = T_0 $,
the gas is rarefied, and it remains an ideal gas in the sense of
relation~\thetag{19}, i.e., in the Boltzmann-Maxwell sense.

There is, however, a fluctuation of the type of a stationary Bose--Einstein
condensate. in this fluctuation, the molecules by themselves placed inside this
fluctuational fragment can possibly move with the same velocities as those of
the gas molecules and, if it were possible to enumerate them, then the numbers
will be changed very quickly. If shall refer to this fluctuation (of the form
of the Bose-Einstein condensation) as liquid, then actual molecules of the
liquid move in it with the same speeds as the gas molecules (of the ``saturated
vapor'').

To represent this picture in a more visible way, imagine a bunting which winds
from one roller to the other. Between the rollers, under the material, a strong
wind blows from a hose. We see a `` hump'' is formed between the rollers;
however, it can be assumed that we do not see that the bunting moves.

Nevertheless, as the density of the Bose condensate increases, our
macroinstrument can fix the bound of the density and show us that there is a
more dense phase and a less dense phase. Hence, only the original
macroinstrument can show us the bound of this abstract liquid, i.e., of the
second phase.

First, the Bose-condensate at rest, i.e., the gas compaction, is being formed
(because of the relativity principle for MRP),
and then there arise quantum forces, i.e., attraction forces (see below),
acting on the ``nearest neighborhoods'',
the more so because the molecules move slower
at a low temperature.

If liquid droplets occur below the temperature of the ``dew point,'' then the
droplets are spherical, even under the presence of the gravity of the Earth
(physicists refer to the very gravity, as a rule, when claiming that the border
between gas and liquid is flat). The pressures in the droplet and in the gas
(the saturated vapor) are different, due to the surface tension.

Therefore, the main rule of the equilibrium ``vapor-liquid,'' namely, the
coincidence of the of pressures, really holds at the dew point only if we
neglect the surface tension, and thus neglect the attraction of liquid
molecules, because these two effects are inseparably linked with each other.
Our concept of a new ideal gas is based on the very assumption on the absence
of attraction between the molecules.

The picture in which the attraction and the surface tension play no role can be
is observed in experiments if the temperature is equal to the temperature $T_0$
of the gas-liquid transition (i.e., at a point of the ``binodal'') and $T_0$ is
still greater than the temperature at which a droplet of critical radius has
been already formed. Then, at $ T = T_0 $, the incipient drops spontaneously
shrink and occur at another point. These drops cannot live without the
surrounding saturated vapor; one can see these drops but cannot feel them.

If the vessel is spherical and the mean free path is comparable to the size of
the vessel (similar to the so-called Knudsen criterion; see ~ \cite{9}--\cite{12},
then the probability of such a virtual drop is larger at the center of the
vessel.

In this case, if we make the labelling of several molecules by launching few
isotopes which can be traced, then these isotopes will pass freely through the
liquid to vapor and back, and they will form a a denser structure near the
center of the ball, in such a way that, when illuminated by parallel rays, it
will provide a shade. However, it is impossible to take this drop from the gas
medium. One can see an ideal liquid but cannot feel it\footnote{One cannot
drink it but can breathe it in.}. Possibly it is better to refer to it as a
``virtual liquid.''

This approach is unusual for the majority of physicists. Although everyone
knows that, say, when photons are collected at a focus at which their
``density'' is high, then it is impossible to separate the focus from the
``photon medium.''

A mathematical analog of the quantum Bose condensate for a classical gas is a
liquid without attraction in which the speeds of the molecules are
approximately the same as the speeds of the molecules in the saturated vapor.
The attraction between molecules results in a significant correction provided
that the radius of a drop is greater than the critical one; however, this
correction abolishes the conditions of the vapor--liquid equilibrium for the
pressure. Therefore, the problem must be divided into two separate problems,
namely, 1) an ideal gas and a perfect liquid without attraction, and 2) the
consideration of the attraction for the case in which the decay into two phases
has already been carried out and the radius of the drop exceeds the critical
value.

\begin{remark}
We define the temperature from the overcondensate part of the system,
i.e., from the gas until the volume of particles in the condensate is comparatively small,
i.e., until the surface tension is formed
and a drop of critical radius size appears.

The drop of critical radius size is the result of a different MR-process,
i.e., of the quantum dipole-dipole interaction clusterization
according to the scheme given in Sec.~4.2.
The nucleation process consists of two mutually related MR-processes.
The Bose condensate in the first MR-process is the nucleation starting mechanism
including the quantum effect of dipole-dipole attraction
and the quantum effect of exchange interaction of identical particles.
\end{remark}

\subsection{Asymptotic continuation of a perfect liquid\\
to the second sheet as the volume of the liquid increases}

In the manual by Landau and Lifshitz and in other manuals, the spectrum is calculated by the
Weyl--Courant formula. Such calculations require the use of the phase volume, and the volume $V$ of
the configuration space naturally arises. We determine the spectrum starting from the number of
degrees of freedom and actually use the volume only in the final result to pass from the number of
particles to the density. As was already seen, the number of degrees is equal to the dimension of
the Bose--Einstein-type distribution.

The gas spinodal, which is defined in a new way as the locus of isotherms of a new ideal gas, is
formed at the maximum entropy at the points at which the chemical potential $\mu$ vanishes.

Therefore, on the diagram $(Z,P_r)$, the spinodal is a segment $P_r\le1$, $Z=Z_c$
in the case of the van der Waals normalization $T_r=T/T_c$ and $P_r=P/P_c$.

Until now we, maximally following the traditional notation used in~[3].
preserve the volume $V$, although \textit{neither the equation for
the $\Omega$-potential given in}~[3, \$ 28]
\begin{equation}
d\Omega=-S\,dT-N\,d\mu \tag{20}
\end{equation}
\textit{nor relations} (1)--(4) \textit{contain the volume $V$}.
We interpret the Bose--Einstein condensate as a liquid phase,
and because for $N>N_c$ the number of overcondensate particles remains constant,
the liquid is ``incompressible''.

For $T_r\le1$, the Bose condensate occurs and, consequently,
for the liquid phase on the spinodal, the quantity
$$
N=T_r^{\gamma_c+1}\zeta(\gamma_c+1)
$$
remains constant on the liquid isotherm. This means that the isotherm of the liquid phase that
corresponds to a temperature $T_r$ is given by
\begin{equation}\label{nor5}
Z= \frac{P_r}{T_r N}=\frac{P_r}{T_r^{\gamma_c+2}\zeta(\gamma_c+1)}. \tag{21}
\end{equation}

All isotherms of the liquid phase (including the critical isotherm at $T_r=1$) pass through the
origin $Z=0$, $P_r=0$ and then fall into the negative region (or to the second sheet). The point
$Z=0$ corresponds to the parameter $\gamma=0$, and hence to the continuation to $\gamma<0$, since,
for $\mu=0$, the pressure
\begin{equation}\label{nor6}
P_r=T_r^{2+\gamma} \frac{\zeta(2+\gamma)}{\zeta(2+\gamma_c)}\tag{22}
\end{equation}
can be continued to $0>\gamma >-1$.

We shall see below that the value of $Z$ as $\mu\to 0$ is also positive, and therefore the spinodal
for $ 0> \gamma> -1 $ gives the second sheet on the diagram $(Z,P)$; it is more convenient to map
this sheet onto the negative quadrant.

Under the assumption that the transition to the liquid phase is not carried out for $T_r=1$, we
equate the chemical potentials $\mu$ and $\tilde{\mu}$ for the ``liquid'' and ``gaseous'' phase on
the isotherm $T_r=1$ (this fact is proved below).

After this, we find the value of the chemical potential corresponding to the transition to the
``liquid'' phase for $T_r<1$ by equating the chemical potentials of the ``liquid'' and ``gaseous''
phases.

In this section, we find the point of the isotherm-isochore of the liquid as the quantity
$\varkappa=-\mu/T$ tends to zero.

First of all, we take into account the fact that $N_c$ is finite, although it is large, and hence we
must use the obtained correction.

In fact, the transition to integral~(6)
from the integral over momenta in~\cite{2}
by using the replacement $p^2/2m=\varepsilon$
corresponds to the transition to the energy oscillatory ``representation''
or, which is the same, to the natural series.
The differential $d\varepsilon$ means that the discrete series
must be taken with the same series in $\varepsilon$, and
this is precisely the natural series multiplied by a small parameter.

Historically, such a representation was present already in the initial Plank distribution. The
transition from the discrete representation of the natural series to the integral representation
will be described in this section. This representation associates the Bose--Einstein distribution
with the number theory considered in Example~1. On the other hand, it stresses that the discrete
Bose--Einstein--Plank distribution depends only on the number of degrees of freedom and is
independent of the three-dimensional volume~$V$.

Obviously, the discrete decompositions leading to integral~(6)
are not unique. Usually, the physicists reduce discrete decompositions to integrals over momenta and
try to relate them to the volume $V$ (and the phase volume, respectively). Using the natural series
and the parameter $\gamma$, we thus stress the difference between these approaches.

Let us construct the thermodynamics of the ideal Bose gas with boundedly many states at a given
quantum level. Since $N_i\leq N$ because of the left equality in formula~\thetag{11-1}, this
condition cannot be an additional restriction. Summing the finite geometric progression, we obtain
\begin{align}\label{v3}
\Omega_i(k)&= \frac{-VT}{\Lambda^{2(1+\gamma)}} \ln\sum_{n=0}^N g_i\left(\exp \left(\frac{\mu}{T}
-\frac{i}{T_r}\right)\right)^n
\nonumber\\
&= \frac{V}{\Lambda^{2(1+\gamma)}}\ln g_i \left(\frac{1-\exp
\left(\frac{\mu}{T} -\frac{i}{T_r}\right)(N+1)} {1-\exp\left(\frac{\mu}{T}
-\frac{i}{T_r}\right)}\right), \qquad g_i= i^{\gamma+1}.\tag{23}
\end{align}

The potential $\Omega$ is equal to the sum $\Omega_i$ over~$i$:
\begin{equation}\label{v4}
\Omega= \sum \Omega_i.\tag{24}
\end{equation}

For the number of particles, we have the formula $N=-\partial \Omega/\partial\mu$
(see (20)). Omitting the
volume~$V$, we obtain
\begin{equation}\label{v5}
N=\frac{1}{\Lambda^{2(1+\gamma)}}\sum_i \left(\frac{i^\gamma}{\exp{(-\frac{\mu}{T}
+\frac{i}{T_r})}-1}- \frac{(N+1) i^\gamma}{\exp{[(N+1)(-\frac{\mu}{T}
+\frac{i}{T_r})]}-1}\right).\tag{25}
\end{equation}

The volume $V$ in relations~\eqref{v3} was required only for the normalization, for the
transition form the number $N$ to the density. For $\gamma>0$, it does not interfere with the
asymptotics as $N\to\infty$, because the term containing $N+1$ in the right-hand side is small. At
the same time, it agrees with the pressure, because $P= -\partial\Omega/\partial V$.

For $\gamma \leq 0$, we omit the volume $V$, because even for $\gamma=0$ due to Example~1, there
appears a term of the form $\ln N$ which must be taken into account\footnote{In this example where
$D=2$ and $\gamma=0$, there is no area $\mathfrak{S}$. And this confuses specialists in
thermodynamics. Indeed, on one hand, $N/\mathfrak{S} \to \text{const}$, but on the other hand, it
follows from~(17) that $\ln M_c \sim 2 \ln N_c$, and hence, by~(17), the limit of
$N/\mathfrak{S}$ as $N_c \to \infty$ and $\mathfrak{S}\to \infty$ tends to infinity. This finally
leads to a false conclusion that the Bose-condensate exists only for $T=0$ in the two-dimensional
case. In fact, it exists for $T_d=\frac{h^2}{\sqrt{2}m}(\frac{N}{\mathfrak{S}})\frac{1}{\ln N}$, and
this is not a very small value (see Corollary~1 below).}, because we have $\ln N \approx 15$ in the
two-dimensional case.

In the two-dimensional trap, the number~$N$ is significantly less, but even for $N=100$, $\ln N=2$,
we can use the asymptotic formulas given below.

On the other hand, the relation between thermodynamic parameters allows us to decrease the number of
independent variables from three to two (cf. Fig.~9 in the variables $\rho, T$ and Figs. 11--16 in
the variables $Z,P$).
\bigskip

\textbf{Estimates}.

$\square$ Taking the parameter~$\gamma$ into account we use the  Euler--Maclaurin formula to obtain
\begin{equation}
\sum_{j}\Big(\frac{j^\gamma}{e^{bj+\varkappa}-1} -\frac{kj^\gamma}{e^{bkj+k\varkappa}}\Big)
=\frac1{\alpha}\int^\infty_0
\Big(\frac{1}{e^{bx+\varkappa}-1}-\frac{k}{e^{bkx+k\varkappa}-1}\Big)\,dx^ \alpha+R,\tag{26}
\end{equation}
where $\alpha=\gamma+1$, $k=N+1$, $b=1/T$, and $\varkappa= -\mu/T$. Here the remainder $R$ satisfies
the estimate
$$
|R|\leq \frac1\alpha\int^\infty_0|f'(x)|\,dx^\alpha, \qquad \text{where}\quad
f(x)=\frac1{e^{bx+\varkappa}-1}-\frac{k}{e^{k(bx+\varkappa)}-1}.
$$
We calculate the derivative and obtain
\begin{align}\label{ad1}
f'(x)&=\frac{bk^2e^{k(bx+\varkappa)}}{(e^{k(bx+\varkappa)}-1)^2}
-\frac{be^{bx+\varkappa}}{(e^{bx+\varkappa}-1)^2},
\nonumber\\[-3\jot]
\tag{27}\\
|R|&\leq \frac{1}{\alpha b^\alpha} \int^\infty_0
\Big|\frac{k^2e^{k(y+\varkappa)}}{(e^{k(y+\varkappa)}-1)^2}
-\frac{e^{y+\varkappa}}{(e^{y+\varkappa}-1)^2}\Big|\,dy^\alpha. \nonumber
\end{align}
We also have
$$
\frac{e^z}{(e^z-1)^2}=\frac1{z^2}+\psi(z),\qquad \text{where $\psi(z)$ is a smooth function and
$|\psi(z)|\leq C(1+|z|)^{-2}$}.
$$
By setting $z=y$ and $z= ky$, we obtain the estimate for~$R$:
\begin{align}\label{ad1a}
|R|&\leq \frac{1}{\alpha b^\alpha} \int^\infty_0
\big|\psi\big(k(y+\varkappa)\big)-\psi(y+\varkappa)\big|\,dy^\alpha
\nonumber\\
&\leq \frac{k^{-\alpha}}{b^\alpha} \int^\infty_{k\varkappa} |\psi(y)|\,dy^\alpha
+\frac{1}{b^\alpha}\int^\infty_{\varkappa}|\psi(y)|\,dy \leq C b^{-\alpha}\tag{28}
\end{align}
with a certain constant~$C$. For example,  if $\varkappa\sim(\ln k)^{-1/4}$, then $|R|$ preserves
the estimate$|R| \sim O(b^{-\alpha})$. $\square$
\bigskip

The energy will be now denoted by~$M$, because without multiplication by the volume~$V$, this is not
the usual thermodynamics but rather a certain analog of the number theory (see Example~1).

Taking account of the fact that, for the value of $M$, the correction in~\eqref{v3} can be neglected
for the value of~$M$, we obtain
\begin{equation}\label{v8}
M= \frac {\Lambda^{\gamma_c-\gamma}}{\alpha\Gamma(\gamma+2)}
\int\frac{\xi\,d\xi^\alpha}{e^{b\xi}-1}=
\frac {\Lambda^{\gamma_c-\gamma}}{b^{1+\alpha}}\int_0^\infty\frac{\eta
d\eta^\alpha}{e^\eta-1},\tag{29}
\end{equation}
where $\alpha=\gamma+1$, $b=1/T_r$. Therefore,
$$
b=\frac1{M^{1/(1+\alpha)}} \left(\frac
{\Lambda^{\gamma_c-\gamma}}{\alpha\Gamma(\gamma+2)}\int_0^\infty\frac{\xi
\,d\xi^\alpha}{e^\xi-1}\right)^{1/(1+\alpha)}.
$$

We obtain (see~\cite{48})
$$
\aligned &\sum_j\Big(\frac{j^\gamma}{e^{bj+\varkappa}-1} -\frac{kj^\gamma}{e^{bkj+k\varkappa}}\Big)
=
 \frac{1}{\alpha}\int_0^\infty
\left\{\frac1{e^{b\xi}-1}-\frac{k}{e^{kb\xi}-1} \right\}\,d\xi^\alpha + O(b^{-\alpha}) \notag
\\
& =\frac{1}{\alpha b^\alpha}\int_0^\infty\left(\frac1{e^\xi-1} -\frac1\xi\right)\,d\xi^\alpha
+\frac1{\alpha b^\alpha}\int_0^\infty\left(\frac1\xi- \frac1{\xi(1+(k/2)\xi)}\right)\,d\xi^\alpha
\notag
\\
&\quad-\frac{k^{1-\alpha}}{\alpha b^\alpha}\int_0^\infty\left\{ \frac{k^\alpha}{e^{k \xi}-1}
-\frac{k ^\alpha}{k \xi(1+(k /2)\xi)}\right\}\,d\xi^\alpha + O(b^{-\alpha})\notag
\\
=&\quad\frac{c(\gamma)}{b^\alpha}(k^{1-\alpha}-1)+O(b^{-\alpha}).
\endaligned
$$
By setting $k=N|_{\tilde{\mu}/T=0}\gg 1$, we finally obtain
\begin{equation}\label{ad2}
N|_{\tilde{\mu}/T=0}\cong (\Lambda^{\gamma_c-\gamma} c(\gamma))^{1/(1+\gamma)}T_r,
\quad\text{where}\quad c(\gamma)=\int_0^\infty\Big(\frac1{\xi}-\frac1{e^\xi-1}\Big)
\xi^\gamma\,d\xi.\tag{30}
\end{equation}

\begin{corollary}{[Erd\H{o}s formula]}
It can be proved that
$\varkappa\to 0$ gives the number $N$ with satisfactory accuracy.
Hence,
$$
N_c=\int_0^\infty\bigg(\frac1{e^{bx}-1} -\frac{N_c}{e^{bN_cx}-1}\bigg)\,dx  +O(b^{-1}).
$$

Consider the value of the integral {\rm(}with the same integrand{\rm)} taken from $\varepsilon$ to
$\infty$ and then pass to the limit as $\varepsilon\to0$. After making the change $bx=\xi$ in the
first term and $bN_cx=\xi$ in the second term, we obtain
\begin{align}\label{13:x300}
N_c&=\frac1b\int_{\varepsilon b}^\infty\frac{\,d\xi}{e^\xi-1} -\int^\infty_{\varepsilon
bN_c}\frac{\,d\xi}{e^\xi-1} +O(b^{-1})=\frac1b
\int^{\varepsilon bN_c}_{\varepsilon b}\frac{\,d\xi}{e^\xi-1}+O(b^{-1})
\tag{31}\\
&\sim \frac1b\int^{\varepsilon bN_c}_{\varepsilon b}\frac{\,d\xi}\xi+O(b^{-1})
=\frac1b\{\ln(\varepsilon bN_c)-\ln(\varepsilon b)\}+O(b^{-1})=\frac 1b\ln N_c+O(b^{-1}).
\tag{32}
\end{align}

On the other hand, making the change $bx=\xi$ in \eqref{v8}, we obtain
$$
\frac1{b^2}\int^\infty_0\frac{\xi \,d\xi}{e^\xi-1}\cong M.
$$
This gives
\begin{equation}\label{15:x300}
b=\bigg({\sqrt M}\bigg/{\sqrt{\int_0^\infty\frac{\xi \,d\xi}{e^\xi-1}}}\,\bigg)^{-1}, \qquad
N_c=\frac12\frac{\sqrt M}{\sqrt{\pi^2/6}}\ln M(1+o(1)) +O(b^{-1}).\tag{33}
\end{equation}

Now let us find the next term of the asymptotics by setting
$$
N_c=c^{-1}M^{1/2}\ln c^{-1}M^{1/2}+\alpha M^{1/2}+o(M^{1/2}), \qquad\text{where}\quad
c=\frac{2\pi}{\sqrt6}\,.
$$
Furthermore, using the formula
$$
N_c=c^{-1}M^{1/2}\ln N_c + O(b^{-1})
$$
and expanding $\ln N_c$ in
$$
\text{ $\frac\alpha{c^{-1}\ln c^{-1}M^{1/2}}\,,$ }
$$
we obtain
$$
\text{ $\alpha=-2\ln\frac c2$. }
$$
Thus, we have obtained the Erd\H{o}s formula~\cite{49}.
\end{corollary}

The relation $N= T_r^{\gamma_c+1}\zeta(\gamma_c+1)$ is consistent with the linear relation
$N=A(\gamma) T_r$, where $A(\gamma)= (\Lambda^{\gamma_c-\gamma} c(\gamma))^{1/(1+\gamma)}$, for
$P_r<0$.

We can normalize the activity $a$ at the point $T_c$, and we can find $a_0$ by matching the liquid
and gaseous branches at $ T_c $ for the pressure $P_r=1$, in order to prevent the phase transition
on the critical isotherm at $T_r=1$.

In what follows, we normalize the activity for $T_r<1$ with respect to the value of~$a_0$ computed
below. Then the chemical potentials (in thermodynamics, the thermodynamic Gibbs potentials for the
liquid and gaseous branches) coincide, and therefore there can be no ``gas--liquid'' phase
transition at $T_r=1$.

Now, for the isochore--isotherm of the ``incompressible liquid'' to take place, we must construct it
with regard to the relation $N_c=\zeta(\gamma_c+1)$, i.e.,
$$
N(T_r) =  T_r^{\gamma_c+1}\zeta(\gamma_c+1).
$$

We obtain the value $\gamma(T_r)$ from the implicit equation
$$
A(\gamma)= T_r^{\gamma_c}\zeta(\gamma_c+1).
$$

Thus, for each $T_r<1$ we find the spinodal curve (i.e., the points at which $\tilde{\mu}=0$) in the
domain of negative $\gamma$~\cite{48},
\begin{equation}\label{1ad}
\Lambda^{(\gamma-\gamma_c)/(1+\gamma)} c(\gamma)^{1/(1+\gamma)} = T_r^{\gamma_c}\zeta(\gamma_c+1),
\tag{34}
\end{equation}

In the set of two values of $\gamma$ corresponding to the solution~\eqref{1ad}, we choose the value
associated with the largest entropy, i.e., the quantity largest in absolute value and denote it by
$\gamma (T_r)$. For $T_r=1$, we choose the value of $\Lambda$ so that both solutions~$\gamma (1)$
coincide, and we write $\gamma_0=\gamma(1)$.

Let $a_g=e^{-\mu/T}$ be the activity of the gas, and let $a_l=e^{-\tilde{\mu}/T}$ be the activity of
the liquid. W e present the condition for the coincidence of~$M$ and of the activities at the point
of the phase transition:
\begin{equation}
T_r^{\gamma_c}\operatorname{Li}_{2+\gamma_c}(a_g)= \Lambda^{|\gamma(T_r)|+\gamma_c}
T_r^{-|\gamma(T_r)|}\operatorname{Li}_{2-|\gamma(T_r)|} \left(\frac{a_l}{a_0} \right),
\tag{35}
\end{equation}
\begin{equation}
\frac{\Lambda^{\gamma_c-\gamma_0}}{\zeta(2+\gamma_c)} \operatorname{Li}_{2+\gamma_0}(a_0)=1, \qquad
a_g=\frac{a_l}{a_0}.
\tag{36}
\end{equation}

\begin{definition}
The relation $a_g=a_l/a_0$ will be called the normalization of activity on the critical isotherm.
\end{definition}

Relations~\thetag{35}--\thetag{36} determine the value of the chemical potential $\mu=\tilde{\mu}=T\ln
a_g$ at which the `` gas--liquid'' phase transition occurs.

Let $T_0=\min_{-1<\gamma<0}A(\gamma)$. Thus, for every $T_0<T<T_c$, we obtain a value of the reduced
activity of the liquid $a_r=a_l/a_0$ ($a_l$ is the activity of the liquid) that corresponds to the
van der Waals normalization.

\begin{remark}
In thermodynamics, the critical values $T_c$, $P_c$, and $\rho_c$ are evaluated experimentally for
almost all gases, and therefore the critical number of degrees of freedom can be set in advance.
According to numerical calculations for a real gas, the parameter $\lambda=1/\Lambda$ ($1.6 <\lambda
<3$, $T_r>1/3$) determining the point $\gamma_0$ ensures that the binodal passes through the triple
point (see Sec.~4.4). The triple point can be determined experimentally with a sufficient accuracy.
\end{remark}

\subsection{Holes in the Bose condensate as observable quantities.\\
The maximum density of holes}

The molecules of an ideal gas can be thought of as tiny balls. Let us imagine
holes, excitons in glass, also as balls which are empty, without the substance
of a  molecule. Obviously, if one mixes these balls in a glass in a chaotic
way, then the chaos in the glass becomes increased. This means that the entropy
increases in the presence of holes. Therefore, to achieve the maximum of the
entropy, we must also additionally mix holes into this glass.

In our conception, holes occur for $ \gamma <0 $.

In the ideal gas model, we ignore the attraction, and this means that, when
``stretching'' the liquid, which results in holes, the liquid does not resist
(as the sand, which is incompressible under the compression and does not resist
under ``tension;'' cf.~the appendix to the book~\cite{16}).

Once there is no attraction, there is no negative pressure `under the
``tension'', i.e., there is no formation of holes. If $\gamma<0$, then the
plane $(Z,P)$ is positive again, and therefore it is covered by the other
sheet. It can readily be seen that the lines entering the point $Z=0$, $P=0$
(i.e., to the point $\gamma=0$) are reflected on this second sheet back, along
the same line. This means that it is geometrically convenient to arrange the
reflection of vectors on the second sheet by using the matrix $-I$, where $I$
stands for the two-dimensional identity matrix, i.e., to flip (carry out the
mirror reflection for) the sheet $\gamma<0 $ to the negative quadrant.

Note that this procedure is compliance with the concepts of Dirac hole theory,
just in the opposite direction, namely, to a hole we assign a negative
pressure, i.e., a negative energy. Now the straight lines can be continued
through the origin to the negative quadrant, although the pressure really does
not change its sign. This is only a convenient geometric ``uniformization.''

Note also that, due to absence of attraction, an ideal liquid is completely
plastic; namely, it does not try to return to the original state (the state
before stretching). In this sense, the Bose condensate for $ \gamma <0 $, which
leads to this ``kind'' of liquid, can also be treated more visually as a glass
or an amorphous solid\footnote{Physicists know that glass is a liquid and an
amorphous metal is a glass. Hence, an amorphous metal is a liquid. It is
probable that the reader will interpret excitons (holes in amorphous metals
and voids in glass) in a simpler way than holes in liquids
because the notion of holes in crystal metals is rather customary.}.
This makes it possible to interpret the state of
the liquid for $ \gamma <0 $ more intuitively.

{\bf Remark~3.} The author has come to the revision of the thermodynamics
when studying economics in which money is the very particles, according the
correspondence principle derived by Irving Fisher. Fisher himself did not
referred to his observation as the correspondence principle. However, since he
was a disciple of Gibbs, there is a clear reason for the fact that the relation
of the basic law of economics
\begin{equation}
PQ=Mv, \tag{37}
\end{equation}
where $Q$ stands for the amount of goods, $M$ for the number of money, $v$ for
the turnover rate, and $P$ for the price of goods, is obviously related to the
correspondence of economical and thermodynamical quantities, namely, the volume
$V$ corresponds to the amount of goods $Q$, the number of money $M$ to the
number of particles $N$, the rate $v$ to the temperature $T$. The price of
goods $P$ is related to pressure  to a lesser extent; however, it is denoted by
the same symbol.

In this correspondence principle, it is natural to correspond holes to debts
and acquitting to annihilation.

As mentioned above, the locus on which the chemical potential is zero gives the
points of maximum entropy. We refer to these points as the ``new spinodal.'' In
economics, this new spinodal means a kind of limit for debts~\cite{15, 17}.
\medskip

Thus, according to the relations thus obtained, we face a double covering of
the plane $\{Z,P\}$ for $\gamma\ge0$ and $-1\le\gamma<0$. The meaning of the
second sheet is that, for $-1\le\gamma<0$, the chaotic state of liquid (as a
phenomenon associated with the Bose condensate) increases when the number of
holes of the type of Frenkel excitons increases, and the holes are placed in
the liquid, which is fluctuationally concentrated on a rather slow-moving
domain ({\it from the point of view of the device discussed above}\footnote{In
reality, the holes can change places with each other and with holes in the
surrounding gas quickly and imperceptibly for the eyes and for the device.}),
in the form of chaotic nanoholes, then the structure of the liquid becomes
chaotically stretched.

Here the holes-excitons cannot be indexed by our device, as well as the
particles, and we can speak only of the density of holes. As was already said
above, it is more convenient to place the second sheet under consideration in
the quadrant $[-Z,-P]$, by continuing the straight lines~\thetag{19} through
the singular point of $Z=0$, $P=0$ to the negative quadrant.  In other words,
to make a reflection with the help of the matrix $-E$, where $E$ stands for the
identity matrix.

Thus, it becomes convenient to speak of `` negative pressure'', although we
neglect the attraction of particles, and hence there can be no negative
pressure at all. As a rule, the pressure, as well as the temperature, is
regarded as a positive quantity. We stretch the liquid, and it becomes
plastically frozen up in this stretched state and does not tend to shrink back.

Let us explain from the point of view of physics why the extension to the
negative square is natural. We compare the new ideal liquid with sand, which is
incompressible under the ``compression'' and ``dost not resist'' under
stretching, because there is no attraction between the grains.
\medskip

{\bf Example 3.} Consider a cylindrical vessel, filled with sand, whose lid
is attached to the piston, in the room of the space station. The increase in
the vessel with the piston leads only to a rearrangement of sand and its
transformation to a floating `` dust'' in the new volume (see~\cite{18}).

If we take into account the gravitational attraction between the grains, then
the phenomenon of pulling the piston creates a negative pressure, and thus it
is natural to pass to the negative quadrant on the $ (P, Z) $ diagram, and then
to neglect the gravitational attraction.
\medskip

Neglecting the presence of attraction here is just as ``legitimate'' as it is
in the theory of vapor-liquid equilibrium, where the condition that the
pressures are equal is possible only if we neglect the surface tension.

This also explains a smooth transition (without a phase discontinuity of the
first kind) of this structure into ice, namely, a frozen glass crystallizes.

\subsection{Critical exponents as observable quantities  under the Wiener
quantization and the derivation of the Maxwell rule}

Mishchenko and the author~\cite{19} considered the transition to a
two-dimensional Lagrangian manifold in the four-dimensional phase space, where
the pressure $P$ and the temperature $T$ (the intensive variables) play the
role of coordinates and the extensive variables (the volume $V$ and the entropy
$S$) play the role of momenta for the Lagrangian manifold, where the entropy is
the action generating the Lagrangian structure.

Seemingly, there is no global canonical transformation leading to a change of
this kind. This does not confuse physicists. For example, in \S 25 of~\cite{2},
``Equilibrium of a solid in an external field,'' it is said that ``from the
equation
\begin{equation}
dE=TdS+\mu dN, \tag{38}
\end{equation}
represented in the form
\begin{equation}
dS=\frac{dE}T-\frac\mu TdN, \tag{39}
\end{equation}
we see \dots''

However, formula~\thetag{38} {\it does not imply} the expression ``represented
in the form''~\thetag{39}. Nevertheless, this ``implies'' the following
conclusion: ``If the field is absent and both $\mu$ and $T$ are constant, then
the pressure is automatically also constant.'' At the same time, the same
textbook states that, at a temperature slightly below the ``dew point,'' ``when
the radius of the drop becomes greater than the critical value, it can be seen
that the pressure of the liquid inside the drop differs from the pressure in
the saturated vapor. The external field is absent. Is this still
thermodynamics? Other words are used; one speaks of a vapor instead of gas and
of the process of nucleation instead of the vapor-liquid equilibrium. And then
a patch is immediately put on the same hole, namely, an extra term is added
to~\thetag{39}. The old thermodynamics has many patches of this kind.

It turns out that this complex transformation, leading to relation~\thetag{39},
can be carried out, as we have seen, only by continuing to the domain of
negative energies. After this, one can justify the Maxwell transition by
introducing a small dissipation (viscosity). The introduction of an
infinitesimal dissipation enables one to simultaneously solve the problem of
critical exponents, without using the scaling hypothesis, on which the method
of renormalization group is based. Let us show this.

In thermodynamics, the viscosity is absent. However, generally speaking,
without an infinitesimal dissipation, an equilibrium in thermodynamics should
not be attained. Therefore, it is natural to implement the occurrence of this
infinitesimal viscosity and then pass to the limit as the viscosity tends to
zero.

The geometric quantization of the Lagrangian manifold (see ~ \cite{20}, \S11.4)
is usually associated with the introduction of the constant $ \hbar $. The
author introduced the term of Wiener (or tunnel) quantization to describe the
case in which the number $\hbar$ is purely imaginary ~ \cite{21, 22}.

Let us apply the Wiener quantization to thermodynamics. The thermodynamic
potential $G=\mu N$ is the action $\mathbb S=\int p\,dq$ on the two-dimensional
Lagrangian manifold $\Lambda^2$ in the four-dimensional phase space
$q_1,q_2,p_1,p_2$, where $q_1$ and $q_2$ are the pressure $P$ and the
temperature $T$, respectively, $p_1$ is equal to the volume $V$, and $p_2$ is
equal to the entropy of $S$ taken with the opposite sign. All other potentials,
namely, the internal energy $E$, the free energy $F$, and the enthalpy $W$ are
the results of projecting the Lagrangian manifold to the coordinate planes
$p_1, p_2$,
\begin{align}
E&=-\int\vec q\,d\vec p, \qquad \vec q=\{q_1,q_2\},\quad \vec p=\{p_1,p_2\},
\nonumber\\
W&=-\int(q_2\,dp_2+q_1\,dp_1), \qquad F=\int(q_1\,dp_1-q_2\,dp_2). \tag{40}
\end{align}

Under the Wiener quantization, we have
$$
N=\varepsilon\frac\partial{\partial\mu}, \qquad
V=\varepsilon\frac\partial{\partial p}, \qquad
S=-\varepsilon\frac\partial{\partial T}.
$$ Consequently, the role of time
$t$ in the quantization, is played by
$\ln(-\mu/T)$,
$$
G=\mu
N\sim\varepsilon\frac\mu
T\frac\partial{\partial(\mu/T)}=\varepsilon\frac\partial{\partial\ln(-\mu/T)}.
$$

Note that the tunnel quantization of the van der Waals equation (vdW) as
$\varepsilon\to0 $ gives Maxwell's rule (see below).

As we shall see below, the critical point and the spinodal point are focal
points. and therefore, as $\varepsilon\to0 $ there points do not come to the
``classical'' picture, i.e., to the van der Waals model. The spinodal points,
which are similar to turning points in quantum mechanics, can be approached by
the Airy function, whereas the critical point, which is the point at which two
turning points are generated (two Airy functions), can be approached by the
Weber function (see~\cite{23}). It is the very Weber function which is used to
express the creation point of the shock wave for $ \varepsilon \to 0 $ in the
Burgers equation is expressed. If one passes to the limit as $ \varepsilon \to
0 $ outside these points, then we obtain the vdW--Maxwell model. However, the
passage to the limit is violated at these very points. Therefore, the so-called
Landau `` classic'' critical exponents~\cite{2} drastically differ from the
experiment. The Weber function give singularities of the form
$\varepsilon^{-1/4}$, whereas the Airy function gives a feature of the form
$\varepsilon^{-1/6}$.

Let us present a more detailed consideration of the Burgers equation.

Consider the heat equation
\begin{equation}
\frac{\partial u}{\partial t}
=\frac\varepsilon2\frac{\partial^2u}{\partial x^2}, \qquad x\in\mathbb R,
\quad t\ge0, \tag{41}
\end{equation}
where $\varepsilon>0$ is a small parameter. As is known, all linear
combinations
\begin{equation}
u=\lambda_1u_1+\lambda_2u_2 \tag{42}
\end{equation}
of solutions $u_1$ and $u_2$
of equation~\thetag{41} are solutions of this equation.

Let us make the change
\begin{equation}
u=\exp(-w(x,t)/\varepsilon).
\tag{43}
\end{equation}
We obtain the following nonlinear equation:
\begin{equation}
\frac{\partial w}{\partial t}+ \frac12\bigg(\frac{\partial w}{\partial
x}\bigg)^2- \frac\varepsilon2\frac{\partial^2w}{\partial x^2}=0,
\tag{44}
\end{equation}
which is referred to as the integrated Burgers equation\footnote{The
usual Burgers equation can be derived from equation~\thetag{44} by
differentiating with respect to $x$ and by using the substitution
$v=\partial w/\partial x $.}. Obviously, to any solution $u_i$ of equation~\thetag{41}
we can assign a solution $ w_i = - \varepsilon \ln u_i $ of the
equation~\thetag{44}, $ i = 1,2 $. To the solution~\thetag{43} of
equation~\thetag{41} we assign a solution
$$
w=-\varepsilon\ln\big(e^{-\frac{w_1+\mu_1}\varepsilon}+e^{-\frac{w_2+\mu_2}\varepsilon}\big)
$$
of the equation~\thetag{34} where $\mu_i=-\varepsilon\ln\lambda_i$, ($i=1.2$). Since
$$
\lim_{\varepsilon\to0}w=\min(w_1,w_2),
$$
we obtain the $(\min,+)$ algebra of the tropical mathematics~\cite{24}.

To find solutions for $ t> t_{\mathrm {cr}} $, Hopf suggested to consider the
Burgers equation
\begin{equation}
\frac{\partial v}{\partial t}+v\frac{\partial v}{\partial
x}-\frac\varepsilon2 \frac{\partial^2v}{\partial x^2}=0, \qquad v|_{t=0}=p_0(x),
\tag{45}
\end{equation}
and to refer to the function $p_{ \text{gen}}=\lim_{\varepsilon\to0}v$ (Riemann
waves) as a (generalized) solution of the equation
\begin{equation}
\frac{\partial p}{\partial t}+p\frac{\partial p}{\partial x}=0, \qquad p|_{t=0}=p_0(x).
\tag{46}
\end{equation}

The solution $v$ of the Burgers equation can be expressed in terms of the
logarithmic derivative
\begin{equation}
v=-\varepsilon\frac\partial{\partial x}\ln u
\tag{47}
\end{equation}
of the solution $u$ of the heat equation
\begin{equation}
\frac{\partial u}{\partial t}=\frac\varepsilon2\frac{\partial^2u}{\partial x^2}, \qquad
u|_{t=0}=\exp\bigg\{-\frac1\varepsilon\int_{-\infty}^xp_0(x)\,dx\bigg\}.
\tag{48}
\end{equation}
Thus, the original problem reduces to the study of the
logarithmic limit of a solution of the heat equation. As is known, the solution
of problem~\thetag{48} is of the form
\begin{equation}
u=(2\pi\varepsilon t)^{-1/2}\int_{-\infty}^\infty
\exp\bigg\{-\bigg((x-\xi)^2+2t\int_{-\infty}^\xi
p_0(\xi)\,d\xi\bigg)\bigg/2th\bigg\}\,d\xi.
\tag{49}
\end{equation}
The asymptotics of the integral~\thetag{49}
can be calculated by the Laplace method. For $ t <t_{\mathrm {cr}}$, we have
\begin{equation}
u=\big(|J|^{-1/2}(\xi(x,t),t)+O(\varepsilon)\big)\exp\bigg\{-\frac1\varepsilon
{\mathcal{S}}(x,t)\bigg\}. \tag{50}
\end{equation}

Here $${\mathcal{S}} (x, t) = \int_{- \infty}^{r (t)} p \, dx ,$$ and the integral
is evaluated along a Lagrangian curve $ \Lambda^t $; $r(x)$ is a point on
$\Lambda^t$. For $t>t_{\mathrm {cr}}$, there are three points $r_1(x)$,
$r_2(x)$, and $r_3(x)$ on $ \Lambda^t $ whose projections to the $x$ axis are
the same; in other words, the equation $ Q (t, \xi) = x $ for $ x \ in (x_1,
x_2) $ has three solutions $ \xi_1 (x, t) $, $ \xi_2 (x, t ) $, and $ \xi_3 (x,t) $.

Write $${\widetilde{\mathcal{S}}}(x, t) = \int_{-\infty}^{r (x)} p \,
dx\quad\text{for}\quad x <x_1,\quad x> x_2,$$ ${\widetilde{\mathcal{S}}}
(x,t)=\min({\mathcal{S}}_1,{\mathcal{S}}_2,{\mathcal{S}}_3)$, and
$${\mathcal{S}}_j = \int_{- \infty}^{r_j (x)} p \, dx,$$ where $J\in\{1,2,3\}$ for $x\in
[x_1, x_2]$.

These arguments enable us to obtain a generalized discontinuous solution
of~\thetag{41} for the times $t>t_{\mathrm {cr}}$. It is defined by a function
$ p = p (x, t) $ defining the significant areas~\cite{21} of the curve ~
$\Lambda^t$. Note that this, in particular, this implies the rule of equal
areas, which is known in hydrodynamics for finding the front of a shock wave
whose evolution is described by equation~\thetag{36}. Note that this precisely
corresponds to the Maxwell rule for the vdW equation.

The solution $ v = v (x, \varepsilon) $ of the Burgers equation at the critical
point $x=p^3$ is evaluated by the formula
\begin{equation}
v(x,\varepsilon)=\varepsilon\frac{\partial\ln u(x)}{\partial x}
=\frac{\int_0^\infty
\exp\{\frac{-x\xi-\xi^4/4}\varepsilon\}\xi\,d\xi}{\int_0^\infty
\exp\{\frac{-x\xi-\xi^4/4}\varepsilon\}d\xi}.
\tag{51}
\end{equation}
As $x\to0$, after the change $\frac\xi{\root4\of\varepsilon}=\eta$, we obtain
\begin{equation}
v(\varepsilon,x)\to_{x\to0}{\root4\of\varepsilon}\cdot \text{const}. \tag{52}
\end{equation}

What does this mean in terms of classical theory and classical measurement,
when the condition referred to in the book~\cite{1} as the `` semiclassic
condition'' is satisfied (i.e., for the case in which we are outside the focal
point)? For the Laplace transform, this means that we are in a domain in which
the Laplace asymptotic method can be applied indeed, i.e., in the domain where
\begin{equation}
u(x)=\frac1{\sqrt\varepsilon}\int_0^\infty e^{-\frac{px-\widetilde
S(p)}\varepsilon}\,dp. \tag{53}
\end{equation}

If the solution of the relation
\begin{equation}
x=\frac{\partial\widetilde{\mathcal{S}}}{\partial p}
\tag{54}
\end{equation}
is nondegenerate, i.e., $\frac{\partial^2\widetilde S}{\partial p^2}\ne0$ at
the point at which $\frac{\partial\widetilde{\mathcal{S}}}{\partial p}=x,$ then the
reduced integral~ \thetag{53} is bounded as $\varepsilon\to0$. For this
integral to have a zero of the order of $\varepsilon^{1/4}$, we must integrate
it with respect to $ x $ after applying the fractional derivative $ D^{-1 /
4}$. The value of $ D^{-1 / 4} $ as applied to 1 (the value of $ D^{-1 / 4} 1$
is approximately equal to $ x^{1/4} $.

By~\cite{25, 47}, the correspondence between the
differentiation operator and a small parameter of the form $D\to1/\varepsilon$
is preserved for the ratio $-\varepsilon\frac{\partial u/\partial x}u$, while
the leading term of the asymptotic behavior is not cancelled in the difference
between $\frac{\partial^2u/\partial x^2}u$ and $\frac{(\partial u/\partial
x)^2}{u^2}$ due to the uncertainty principle (see Remark~4).
\medskip

{\bf Remark~4.} Let us repeat the calculations in~\cite{1} with regard to the
fact that, on this class of functions, $\overline {D} $ has the properties
$\int\varphi D\varphi\,dx=\frac12\int D\varphi^2\,dx=0$ and $\int
x\varphi^2\,dx=0.$

Consider the obvious inequality
\begin{equation}
\int_{-\infty}^{+\infty}\bigg|ax\psi+\frac{d\psi}{dx}\bigg|^2\,dx\ge0, \tag{55}
\end{equation}
where $a$ is an arbitrary real constant. When evaluating this integral, we have
$$
\int x^2|\psi|^2\,dx=\overline{(\Delta x)^2},
$$
$$
\int\bigg(x\frac{d\psi^*}{dx}\psi+x\psi^*\frac{d\psi}{dx}\bigg)\,dx=
\int x\frac{d|\psi|^2}{dx}\,dx=-\int|\psi|^2\,dx=-1,
$$
\begin{equation}
\int\frac{d\psi^*}{dx}\frac{d\psi}{dx}\,dx=
-\int\psi^*\frac{d^2\psi}{dx^2}\,dx=
\frac1{\varepsilon^2}\int\psi^*|D|^2\psi\,dx=
\frac1{\varepsilon^2}\overline{|\Delta D|^2}.
\tag{56}
\end{equation}
We obtain
\begin{equation}
a^2\overline{(\Delta x)^2}-a+\frac1{\varepsilon^2}\overline{|\Delta D|^2}\ge0.
\tag{57}
\end{equation}
For this quadratic trinomial (in $a$) to be positive for all values of $a$, it
is necessary that the following condition be satisfied:
$$
4\overline{(\Delta x)^2}\frac1{\varepsilon^2}\overline{|\Delta D|^2}\ge1
$$
or
\begin{equation}
\sqrt{\overline{(\Delta x)^2}\ \overline{|\Delta D|^2}}\ge\frac\varepsilon2.
\tag{58}
\end{equation}
Thus, the tunnel quantization explains both the condition $\mu = 0 $ for
photons and the condition $\mu\leq 0 $ for bosons.
\medskip

In the case of thermodynamics, the role of $x$ is played by the pressure $P$,
and the role of the momentum $p$ is the played by the volume $ V $. Therefore,
$ V \sim P^{1/4}$, i.e.,
\begin{equation}
P_c\sim(V-V_c)^4.
\tag{59}
\end{equation}

This is the very jump of the critical exponent. One van similarly obtain other
critical exponents (see~\cite{25}). For the comparison with experimental data,
see the same paper.

Unfortunately, thermodynamics does not use the concept of Lagrangian manifold
which was introduced by the author in 1965~\cite{26}. It is especially suitable
for thermodynamics, in which there are pairs of intensive and extensive
quantities. Intensive quantities, roughly speaking, are the quantities for
which one cannot create the concept of ``specific'' quantity. These are the
temperature $T$, the pressure $P$, and the chemical potential $\mu$. To these
intensive quantities, there correspond related extensive quantities, namely,
the entropy $S$, the volume $V$, and the number of particles $N$. Altogether,
they form the phase space, where the role of coordinates is played by the
intensive quantities and the role of momenta is played by the extensive
quantities. In this case, a Lagrangian manifold is a three-dimensional
submanifold (of the six-dimensional phase space) on which there is an action,
an analog of the integral ${\mathcal{S}}=\int p\,dq,\qquad q\in \mathbb{R}^2$,
$p\in\mathbb{R}^2$ in mechanics. It is locally independent of the path.

Usually a 4-dimensional phase space $ T, S; P, V $ is considered. This space
corresponds to $q\in R^2,p\in R^2$, $q_1\to T$, $q_2\to P$, $p_1\to S$, $p_2\to
V$: $d{\mathcal{S}}=p_1\,dq_1+p_2\,dq_2,$ depending on the coordinate plane of the
form $q_1,q_2$; $q_1,p_2$; $p_1,q_2$; $p_1,p_2,$ to which the Lagrangian
manifold is projected.

The Lagrange property means that the number of planes cannot coincide (there
are no planes of the form $ q_1, p_1 $ and $ q_2, p_2 $). To every projection
there corresponds some potential ($q_1, q_2 $ is the thermodynamic potential,
etc.).

This is an obvious correspondence. If it were more elaborated, then, on one
hand, the transition from the action $d{\mathcal{S}}=p_1\,dq_1+p_2\,dq_2$ to the
``action''--coordinate $dq_1=d{\mathcal{S}}/p_1-(p_2/p_1)\,dq_2$ would be not ``so
obvious'' (see formulas~\thetag{28}--\thetag{29}). On the other hand, it would
be natural to use the semiclassical (Wiener--Feynman) quantization of action rather than
the scaling hypothesis.

(The quantization of the Lagrangian manifold differs from the full quantization
of thermodynamics~\cite{27} in the same way in which the semiclassical
Bose--Sommerfeld geometric quantization differs from the quantization of
Schr\"odinger, Heisenberg, and Feynman.

The term ``dequantization,'' which is well-known in the tropical
mathematics~\cite{24}, means the Wiener or tunnel quantization of the
Lagrangian manifold, and then the passage to the limit as the quantization
parameter (the viscosity) tends to zero.

\section{Nuclear physics in nano scale}

\subsection{Rotation of a neutron in the coat of Helium-5}

 In~\cite{RJMP_24-3},~\cite{RJMP_24-4}, the author
introduced a hidden parameter
$t_{\mathrm{meas}}$
(measurement time)
binding together quantum
and
classical mechanics.
 The author
considered
this parameter
using helium-4, helium-5,
and
helium-6 as examples.
 A detailed proof of the theorem
involving the hidden parameter
for
helium-5
invokes a considerable number of auxiliary statements
and theorems.
 The author,
essentially,
proved
and
discussed
all these auxiliary statements,
such as
approximations based on the Hartree equation
in the case of the Bose distribution
in his earlier papers
(see~\cite{Masl_Feinm_Integral},~\cite{29}).
 In particular, the author
obtained a rigorous correction to the Stefan--Boltzmann law~\cite{Steph-Bolt-TMF}
and proved that the formal series
defining the succeeding terms are false.
 In Gentile statistics (parastatistics)~\cite{Gentile},
 the author
also obtained
a number of estimates
and lemmas
(see
for example,~\cite{MTN_100-1}--\cite{MTN_102-2}).

 In the present Section,
we present only
a scheme of proof of the applicability of the hidden parameter
$t_{\mathrm{meas}}$
introduced by the author~\cite{MTN_102-6_sh_com}
for explaining the behavior of the neutron
in the coat of helium-5.
 The detailed proof
is contained
in the asymptotics obtained by the author
earlier.
 Here we shall present the material
in such a way as, on the one hand, to make it accessible to
mathematicians who studied the author's papers
and,
on the other hand,
to make it clear to nuclear physicists.

 The parameter under discussion
in this paper is not
hidden
in the sense that was attributed
to it by the authors of the Einstein--Podolsky--Rosen paradox (EPR).
 This parameter
is a completely natural
and clear parameter.
 In the quotation from the book~\cite{1} dealing with
the identity of the particles
often referred to in the author's papers,
this parameter is veiled:
this is the ``instant of time,''
at which the numbering of particles
is achieved.
 They wrote:
``the particles belonging
to a given physical system can be considered as `numbered'
at some instant of time~\cite[p.~252 of the Russian edition]{1}.''
 It is   the time during which the particles were numbered
 that   was introduced as an additional parameter
in~\cite{RJMP_24-3},~\cite{RJMP_24-4},~\cite{MTN_102-6_sh_com}
and which is considered
in the present paper.
 This time
depends
on the algorithm used for the numbering of particles.
 The time
needed for the operation of the algorithm,
in turn, depends
on the computing facilities.
 Thus,
this parameter
is not
hidden, but is veiled; it can be determined exactly only under
a large number of additional conditions.

\noindent\textbf{1.} The self-consistent equations obtained by the author,
were first given
in~\cite{15}.
 They relate the Gentile statistics  with  the Bose-Einstein statistics and the Fermi-Dirac statistics:
\begin{align}
\label{eg-N}
 N&=
 V T^{\gamma+1}(\operatorname{Li}_{1+\gamma}(a)-\frac{1}{(N^\alpha+
1)^{\gamma}}\operatorname{Li}_{1+\gamma}(a^{N^\alpha+1})), \tag{60}
\\*
\label{eg-M}
 M &=VT^{\gamma+2}(\operatorname{Li}_{2+\gamma}(a)-\frac{1}{(N^\alpha+1)^{1+
\gamma}}\operatorname{Li}_{2+\gamma}(a^{N^\alpha+1})), \tag{61}
\end{align}
where the
$M=\Omega$
is the potential,
$\Omega = -VP$,
$P$
is the pressure,
$V$ is the volume,
$T$ is the temperature,
$a$
is the activity, and
$D=2\gamma +2$
is the number of degrees of freedom.
 The value of~$\alpha$
varies
from~$1$
to~$0$.

 The case in which
we set
$N\log(a)\to 0$
in the macroscopic equations~\eqref{eg-N}--\eqref{eg-M},
belongs
to
mesoscopic physics.
 In
thermodynamics,
this situation
arises
near the point of change of the sign of the activity~$a$,
i.e.,
near the point of passage of the Bose--Einstein distribution
to the Fermi--Dirac distribution
(for
$\gamma \leq 0$)%
\footnote{This passage
was studied in
great detail
by the author
in the theory of decomposition of rational numbers;
see,
in particular, the paper~\cite{MTN_102-2},
where the author sews together the boson and fermion branches.
Further,
mesoscopy arises
between the values
$N=0$
and
$N=1/\log(a)$
according to abstract analytic number theory
(for a detailed bibliography on
analytic number theory,
see the book~\cite{Postnikov}).}.

\begin{remark}
 It is well known that
thermodynamics
can be carried over
to economics.
 Thus, Irving Fisher
associated the price of goods with the quantity~$PV$
and the amount of money with the number of particles~$N$.
 The author
associated the amount of debts with the negative values of~$N$.
 In
economics,
$N=0$
or
$N<0$
are regarded as a day-to-day usual situation.
\end{remark}

 We introduce the notation
$W= V (\lambda^2 T)^{\gamma+1}$,
and let
$\lambda$
be the parameter
depending
on the mass.
 In
number theory,
$\gamma=0$ and
$\lambda=1$~\cite{MTN_102-2}.

 Here we apply Gentile statistics (parastatistics)
and relate it, in a self-consistent way,
to the statistics of bosons and fermions
in mesoscopic physics.

\begin{figure}[h!]
\begin{center}
\includegraphics{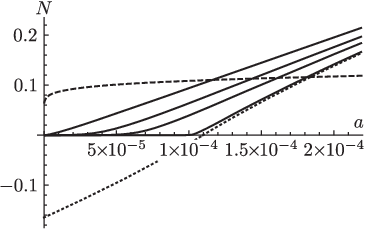}
\end{center}
\caption{Numerically found exact dependence of~$N$
on~$a$.
 The solid curves
correspond to
$\gamma=0$,
$\lambda=1$,
$W=1000$,
$\alpha=0.5, 0.8, 0.9, 0.9999$
(from left to right).
 The points
correspond to
$\alpha=1$
because of two reflections.
 The dotted curve
corresponds to
$N= -1/\log (a)$.}
\end{figure}

 Consider the case in which
$N_i$
is the number of holes.
 This means that
we assume
$N_i$
to be a negative number.
 Then
\begin{equation}
\label{24-4-1}
\sum(-N_i) = -N, \qquad
-\sum \varepsilon_i N_i =-M.\tag{62}
\end{equation}
 Thus, the numbers
$-N$
and
$-M$
are also
negative.
 The multiplication
of both equalities
by
-1 leads to the same case
in which
the
$N_i$
are positive.
 Therefore,
the formulas of Gentile statistics
remain the same.
 This means that, in the formulas
of Gentile statistics,
we can replace the numbers
$N$
and
$M$
by their absolute values.
 In this way,
we can extend Gentile statistics
to negative numbers~$N_i$,
i.e.,
to the case of holes.

 To extend the self-consistent formulas
of the statistics
introduced
by the author
in~\cite{15}
to the case of ``holes,''
we must extend
the curves given in Fig.~6A
into the negative domain,
using their mirror reflection in the axis~$a$
(see Fig.~7A).

\begin{remark}
 If we use the mirror reflection of the appropriate curve
with respect to the line
$a=a_0$,
where
$a_0$
satisfies the equation
$N(a_0)=0$,
then
the curve
will be
a continuously differentiable continuation of the original curve
describing the self-consistent equation
for
$N>0$.
 The second derivative
undergoes a jump.
\end{remark}

\begin{definition}
 The distance from the point
$a=0.00010$,
where
$\alpha=0$,
$N=0$,
to the intersection point
with the curve
$N=1/\log(a)$
is called the \textit{spin concentration}.
\end{definition}

 This concentration
plays a role similar to~$\Delta m$
in Einstein's formula $E=\Delta m c^2$,
 where  $c$  is  the velocity of light ~\cite{MTN_102-6}.

 Our expansion
in the small parameter
$N \log(a)$
will bound
the curve described above by its intersection point
with the line
$N=1/log(a)$.
 At this point,
as is seen from
Fig.~6A, the
$\Omega$-potential
attains
its largest absolute value on the closed interval
$|N|= |1/\log(a)|$.

\begin{figure}[h!]
\begin{center}
\includegraphics{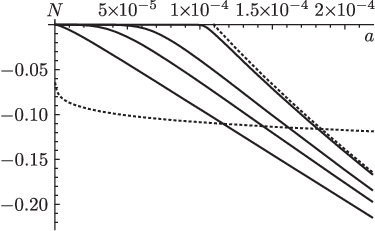}
\end{center}
\caption{The reflection of the curves $N(a)$
depicted in Fig.~6A
with respect to the line
$N=0$.
 The solid curves
correspond to
$\gamma=0$,
 $\lambda=1$,
$W=1000$,
$\alpha=0.5, 0.8, 0.9, 0.9999$
(from left to right).
 The points
correspond to
$\alpha=1$
in view of two reflections.
 The dotted curve
corresponds to
$N= -1/\log (a)$.}
\end{figure}

 In the positive domain of energy,
the curves
$\alpha = 0.5, 0.8, 0.9, 1$
are symmetric
with respect to the axis~$a$; hence
a similar jump in the energy
occurs
also in the positive domain of energy.
 Thus, particles from the positive domain jump into
the symmetric negative domain.

 To the author's knowledge, such an energy jump in a transition
occurs in thermodynamics
only
in the quantum case
and
in the case of a capillary with superfluid helium-4
at the point at which
the Allen--Jones spouting occurs~\cite{FAN-2003}.
 Therefore,
in
our case,
we can assume that, at the point of passage of the Einstein--Bose distribution
to the Fermi--Dirac distribution
a similar ``spouting''
on a mesoscopic scale occurs.
 In
our case,
 the Allen--Jones ``spouting''
is the phenomenon in which
one neutron breaks away
and
goes to infinity
with
the velocity
obtained in the energy jump.

 Let us pass
to the model of the helium nucleus.
 According to Bohr, the nucleons inside the shell of the nucleus
do not interact
(there is no attraction between them).
 They act as colliding balls.
 Indeed,
according to the latest experiments, nucleons
attract to one another only
at distances less than or equal to their radii.

 But
this  fact is also an approximation.
 Indeed, by the Schr\"odinger equation, the nucleon
is a wave packet.
 Therefore, it
spreads from its original
$\delta$-shaped structure and, therefore, there is a small interaction
$\varepsilon V(x-y)$
between nucleons.
 Here
$\varepsilon$
is a small parameter,
$x$
corresponds to
one nucleon and
$y$
to the other,
and
$V(x-y)$
is
the interaction potential.
 The parameter
$\varepsilon$
is a `` hidden'' parameter.
 For
$\gamma>0$,
the number of degrees of freedom depends
on the relationship between this parameter
and the Planck constant.

 This implies the following:

(1) the   maximum  number of degrees of freedom
of the nucleon
is
$6n-5$,
where
$n$
is the number of nucleons
(\cite[Sec.~44]{2});

(2) by the self-consistent Hartree equation
for fermions and bosons,
the interaction potential
$\varepsilon V(x-y)$
for
helium-4
constitutes a double shell---the first shell
with a high barrier and the second shell
with a low barrier.
 The first shell
contains
two neutrons
and two protons,
while the second shell
contains
two neutrons for at most 1\,s,
forming helium-6.
 We can
state conditionally that,
outside this time interval,
between
the main (first) shell
and the second shell,
there are
two ``holes,''
which, occasionally, are filled by neutrons.

 The distance
between the two shells
constitutes the so-called \textit{coat}.
 The given construction
refines the initial Bohr model
in which the nucleons
do not attract one another.

We obtained  the following relation for~$W=V(\lambda^2T)^{\gamma+1}$
$$
 W=\frac{E_{\mathrm{nuc.bin}}}{T_c \zeta({2+\gamma})(\gamma+1)},
$$
where $E_{\mathrm{nuc.bin}}$   is the nuclear binding energy.
 For helium-4,
the value of~$W$
turns out to be
$3.1\times10^{10}$.
 At the intersection point of the graphs
$N=-1/\log(a)$
and
$N(a)$
(see~Fig. 8A), the energy
$E$
is
$0.12 \ T_c$
or
$53.1\times10^{-6}$\,eV .

\begin{figure}[h!]
\begin{center}
\includegraphics[draft=false]{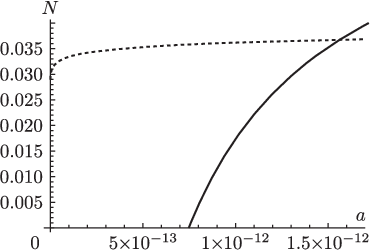}
\end{center}
\caption{The solid line
corresponds to the dependence
$N(a)$;
the dashed line describes the function
$N=-1/\log(a)$.
 Here
$\gamma=0.5$,
$\lambda=1$
$W=4.7\cdot 10^{10}$.}
\end{figure}

Hence,
by defining
$$
t_{\mathrm{meas}}=\hbar/E_{\mathrm{ms}},
$$
we obtain
  $t_{\mathrm{meas}}=1.24\times10^{-11}$\,s.
  For  $\gamma=2.5$,  the mesoscopy (the number of particles is less than $10^6$)
passes into a microscopy (the number of particles is~2).

   In the above-mentioned paragraph from the book~\cite{1},
 Landau and Lifshits wrote about another instants of time:
if ``in the following time,
one observes the motion of each particle
along its own trajectory,
then,  at any
\emph{instant of time}
(italicized by me---VM), the particles
can be identified.''

 Note that the time
during which the experimenter sees the behavior of the particles
is much less than the time
$t_{\mathrm{meas}}$
of the experiment
(of the numbering)
(and at least
100 times less than the lifetime
of the fermion of helium-5).
 Instants of time
constitute a discrete collection of points.
 If the time intervals
between these points are much less than
the veiled parameter,
then the observer
will see the classical pattern of rotation of the neutron
(of the wave packet)
about the nucleus of helium-4
regardless of the lifetime
of the fermion of helium-5.

\subsection{Nuclear decay}

The development of wave mechanics started from de Broglie's paper  ``Ondes et quanta'' in 1923.
De Broglie considered the motion of electron in a closed orbit and showed that
the requirement that the phases be consistent results in the Bohr--Sommerfeld quantum condition,
i.e., to the quantization of the angular momentum.
In 1927, developing his ideas about the relation between waves and particles,
de Broglie constructed the \emph{theory of double solution}~\cite{de_Broglie}
which, in fact, resulted in the well-known notion of wave-corpuscle dualism, which is still actual nowadays.

De Broglie concluded that the presence of a continuous wave is related to the fact
that the Lagrangian of the particle contains an additional term which can be treated
as a small addition of the potential energy (cf. formula~\eqref{N01-0} below).
This theory agrees well with the so-called Bell inequalities~\cite{Bell} and is a nonlocal theory.

\bigskip

\textbf{a. {Bose statistics and Fermi statistics in the Hougen--Watson diagrams and in the Gentile statistics}}

Bohr and Kalckar investigated the boson nucleus~\cite{Bohr_Kal}.
The capture of one neutron turns the boson nucleus into a fermion.

The behavior of Bose and Fermi particles is described by the Bose--Einstein and Fermi--Dirac distributions, respectively.
The Bose--Einstein distribution in polylogarithm form becomes
 \begin{equation}\label{B-E}
    \operatorname{Li}_s(a) = \frac{1}{\Gamma(s)} \int_0^\infty \frac{t^{s-1}}{e^t/a-1}\, dt, \tag{63}
      \end{equation}
where $\operatorname{Li}_{(\cdot)}(\cdot)$ is a function of the polylogarithm.
The Fermi--Dirac distribution can be written as
  \begin{equation}\label{F-D}
   - \operatorname{Li}_s(-a) = \frac{1}{\Gamma(s)} \int_0^\infty \frac{t^{s-1}}{e^t/a+1}\, dt.\tag{64}
      \end{equation}

We consider the quantum particles each of which is associated with a wave packet.
These wave packets are related to the de Broglie thermal wavelength~$\Lambda$.

Assume that $a =e^{\mu/T}$ is the activity ($\mu$ is the chemical potential),
$s=D/2$, and $D$ is the number of degrees of freedom (dimension). We denote the total energy
of all $N$ particles (molecules) by~$E$.

One can see that, for the activity $a$ changing sign,  the distributions~\eqref{B-E} and~\eqref{F-D} themselves
also change sign. This corresponds to the transition from negative pressures to positive pressures.
In the van der Waals formulas~\cite{MTN_97-6},~\cite{RJMP_22-3}, such a picture is rather natural.
Thus, the fermions and bosons are located in different quarter on the Hougen--Watson PZ-diagram
($P$ is the pressure,  $Z=PV/(NT)$ is the compressibility factor,  where $V$ is the volume,
$N$ is the number of particles, and $T$ is the temperature):
the bosons are in the negative domain and the fermions are in the positive domain.
Developing the Bohr--Kalckar approach to the relation between the nucleus model
and formulas of decomposition of an integer into terms
(see the work~\cite{Bohr_Kal} mentioned above), the author showed that this phenomenon
is also manifested in diagrams of the number theory~\cite{RJMP_25-2}.

\bigskip

\textbf{b. {Gentile statistics}}

In physics, the Bose--Einstein and Fermi--Dirac distributions are determined by using the Gentile statistics~\cite{Gentile}.
The Gentile statistics comprises the Bose statistics and the Fermi statistics are particular cases.
The Gentile statistics contains an additional constant~$K$ which denotes the maximal number of particles
located at a fixed energy level. In particular, for $K=1$, the distributions of the Gentile statistics
coincide with the distributions of the Fermi--Dirac statistics,
i.e., the formulas coincide with~\eqref{F-D} in form.
In the Gentile statistics, one assumes that $K\geq1$.

Considering the $\Omega$-potential  corresponding to the Gentile statistics,
we can obtain a detailed description of the boson-to-fermion transition. And judging by analogy with the $\Omega$-potential
considered by Landau and Lifshits~\cite{2}, this allows one to calculate the energy of this transition.
In~\cite{2}, for the case $s=3/2$, the following formula for the total energy of gas is given:
\begin{equation}\label{TotEn}
E=\int_0^\infty \varepsilon \, d N_\varepsilon=
\frac{gVm^{3/2}}{\sqrt{2}\pi^2\hbar^3} \int_0^\infty \frac{\varepsilon^{3/2}\, d\varepsilon}{e^{(\varepsilon-\mu)/T}-1}.\tag{65}
\end{equation}

The fermions in the boson are ``experimentally'' indistinguishable:
if two fermions constituting a boson are interchanged, then these states cannot be distinguished experimentally.
If a boson splits into two fermions that can be distinguished,  they cannot interchange their places and transform into each other.
One can say that the fermions are experimentally distinguishable or indistinguishable
depending on whether the experiments permit distinguishing the fermions comprising the boson.

Our goal is to determine the total energy of transition
of a boson consisting of two experimentally indistinguishable fermions
into two  distinguishable fermions.
This transition occurs in the following two stages:
the transition from indistinguishable fermions (boson) to the distinguishable fermions
and then the disappearance of a fermion. As a result, the processes reduces to the transition
of formula~\eqref{B-E} into formula~\eqref{F-D}.

\bigskip
\textbf{c. {Notation}}

Let us introduce the new notation which permits determining the energy in dimensionless form.

Let $\mathfrak{v}=\Lambda^{2s}$. This quantity has the dimension of volume in the $2s$-dimensional space.
Let $\mathbf{E}=\frac{2\pi\hbar^2}{m}{V}^{-\frac{1}{s}}$. This quantity has the dimension of energy .

Now we introduce dimensionless variables, $\mathfrak{E}={E}/{\mathbf{E}}$ for the total energy
and $\mathfrak{V}={V}/{\mathfrak{v}}$ for the volume.
We note that the quantity $\mathfrak{V}^{1/D}$ is the ratio of the characteristic linear dimension
of the system ${V}^{1/D}$ to the de Broglie wavelength~$\Lambda$.

Usually, $N_i$ denotes the number of particles located at the $i$th energy level.
It is assumed that, in the case of the Fermi gas, there is at most one particle at each energy level,
and in the case of the Bose gas, the number of particles~$N_i$ at each energy level can be arbitrarily large.
We consider the Gentile statistics~\cite{Gentile} according to which, at each energy level,
the number of particles located at each energy level is bounded by the number~$K$.
In other words, the number of particles at any energy level cannot exceed the number~$K$.

The maximal number of particles at an energy level in the system
is attained for the maximal value of the activity~$a$, i.e., at the point $a=1$.
Since $\sum_{i=1}^M N_i = N$, it is obvious that $N_i\leq N$ for the Bose system.
Therefore, $K\leq N$ for the Bose system.
In the Gentile statistics, the~$K$ are integers such that $K_i<K_{i+1}$.

\bigskip
We assume that $K=N$ in an infinitely small neighborhood of $[N]$, where $[N]$ is the integral part of the number~$N$.

In the nonstandard analysis developed by Robinson (see~\cite{Nestandart-1}--\cite{Nestandart-2}),
the set of points infinitely close to the number $[N]$ is called the Leibnitz differential~\cite{Shepin-2}
which is understood as the length of an elementary infinitely small interval (monad).
The differential is an arbitrary infinitely small increment of a variable.

By $x$ we denote the difference $N-[N]$, i.e., $N-[N]=x>0$.
We seek the expansion in a power series in~$x$ up to $O(x^2)$, which implies that $N\sim [N]$.

For the ideal gas of dimension~$D$ obeying the Gentile statistics,
i.e., in the case where, at each energy level, there can be at most~$K$ particles
($K$ is an integer), the following relation for the number of particles~$N$ is known:
\begin{equation}\label{Gent}
N=\mathfrak{V}(\operatorname{Li}_{s}(a) -\frac{1}{(K+1)^{s-1}}\operatorname{Li}_{s}(a^{K+1})) .
\tag{66}
\end{equation}

The self-consistent relation for $x$ in a neighborhood of $[N]$ has the form
\begin{equation}\label{Np}
[N]+x=\mathfrak{V}(\operatorname{Li}_{s}(a)-\frac{1}{([N]+x+1)^{s-1}}\operatorname{Li}_{s}(a^{[N]+x+1})).
\tag{67}
\end{equation}

The following thermodynamical formula for the energy is known:
\begin{equation}\label{Mp}
\mathfrak{E}= s \mathfrak{V}^{\frac{s+1}{s}}(\operatorname{Li}_{s+1}(a)-\frac{1}{([N]+x+1)^s}\operatorname{Li}_{s+1}(a^{[N]+x+1})).
\tag{68}
\end{equation}

We note that, in the thermodynamics, $N$ is the number of molecules. In this paper, we do not consider molecules,
we only consider the nucleus, i.e., the nuclear physics. In this sense, we can speak that, in our model,
the number of molecules $N$ is zero. Therefore, in contrast to the standard Gentile statistics,
we also assume that $K=0$, and we consider only the case $[N]=0$. To the numbers $N=K=0$ we apply the nonstandard analysis
and the technique of the Gentile statistics~\cite{Gentile}.

Using the technique of nonstandard analysis, we add a monad~$x$ to the integer~$K$.
Then expression~\eqref{Gent} is not equal to zero.

We expand the right-hand side of Eq.~\eqref{Np} in small $x\neq 0$ omitting the third-order terms:
\begin{equation}     \label{N01-0}
\begin{split}
&x=\mathfrak{V} x ((s-1) \operatorname{Li}_{s}(a)-\log (a)\operatorname{Li}_{s-1}(a))\\
&
+\mathfrak{V} \frac{1}{2} x^2 \left(\log ^2(a) (-\operatorname{Li}_{s-2}(a))-(s-1)
(s \operatorname{Li}_{s}(a)-2 \log (a)\operatorname{Li}_{s-1 }(a))\right),
\end{split}
\tag{69}
\end{equation}

Cancelling $x$ in both sides of~\eqref{N01-0} and passing to the limit in~\eqref{N01-0},  $x\to0$,
we obtain an expression for~$a_0$, i.e., the value of~$a$ at which $N=0$:
\begin{equation}
\label{N=0}
(s-1) \operatorname{Li}_{s}(a_0)-  \log (a_0)\operatorname{Li}_{s-1}(a_0)-\mathfrak{V}^{-1}=0.
\tag{70}
\end{equation}

The value $ \operatorname{Li}_{s}(a)$,  where $a=e^{\mu/T}$, is associated with the total energy of transition,
in particular, in the three-dimensional case ($s=3/2$).

We note that it follows from Eq.~\eqref{N=0} that, $a_0\to 0$ as $\mathfrak{V}\to{\infty}$.
This means that the values~$a_0$ are small in the case where the value of the system characteristics linear dimension,
which is equal to~${V}^{1/D}$, exceed the de Broglie thermal wavelength~$\Lambda$.

For a sufficiently large value $\mathfrak{V} =\frac{V}{\Lambda^{2s}}$, Eq.~\eqref{N=0} has a unique solution
$a_0\le1$ which depends on $\frac{V}{\Lambda^{2s}},s$. We have
\begin{equation}
\label{N=02}
(s-1) \operatorname{Li}_{s}(a_0)-  \log (a_0)\operatorname{Li}_{s-1}(a_0)=\frac{\Lambda^{2s}}{V}.
\tag{71}
\end{equation}

The expression for the de Broglie thermal wavelength $\Lambda$ has the form
$\Lambda=\sqrt{\frac{2\pi\hbar^2}{m T}}$.

The value of the activity $a$ at a known temperature~$T$ determines the following value of the chemical potential~$\mu$:
\begin{equation}
\label{mu0}
\mu=T \log(a)\le0.
\tag{72}
\end{equation}

In particular, at $a=a_0$, the greater the temperature $T$, the less $a_0$ and the greater the corresponding value $|\mu_0|$.
Thus, as the temperature increases, the transition point~$\mu_0$ approaches the point $\mu=-\infty$
at which the pressure~$P$ changes sign.

Assume that $a_0=1$ and the mass $m$ and the volume $V$ of the nucleus are known.
Then, taking the expression for the de Broglie thermal wavelength $\Lambda=\sqrt{\frac{2\pi\hbar^2}{m T}}$ into account,
we can consider Eq.~\eqref{N=02} as an equation for~$T$.

The temperature arising at $a=1$, i.e., as $\mu \to 0$, will be called the critical temperature.
We denote it by~$T_s$.
Since the temperature $T_s$ is the lowest on the whole interval of variation in~$\mu$
which is the ray $(-\infty,0]$, the ratio $T/T_s$ will be called the regularized temperature,
and we denote it by $T_{\text{reg}}$. The temperature variation along the isotherm can be measured
in~$T_{\text{reg}}$.

The expansion of the energy~\eqref{Mp} in small~$x$ up to the first order inclusively has the form
\begin{equation}
\label{E1}
\mathfrak{E}=s \mathfrak{V}^{\frac{s+1}{s}} x (s \text{Li}_{s+1}(a)-\log (a) \text{Li}_{s}(a)).
\tag{73}
\end{equation}

The ratio of the total energy $\mathfrak{E}$ to the number~$x$
will be called the nonstandard specific energy.
Let us calculate the nonstandard specific energy at the point $a_0$ of boson-to-fermion transition.

Thus, at the point $a=a_0(\mathfrak{V},s)$, the values of the  nonstandard specific energy $\mathfrak{E}_{sp0}$ and ${E}_{sp0}$
are expressed by the formulas
\begin{equation}
\label{Esp0bb}
\mathfrak{E}_{sp0}(\mathfrak{V},s)=s \mathfrak{V}^{1+1/s} ( s\operatorname{Li}_{s+1}(a_0)-\log (a_0) \operatorname{Li}_{s}(a_0)),
\tag{74}
\end{equation}
\begin{equation}
\label{Esp0}
{E}_{sp0}(\mathfrak{V},s,T)=s\mathfrak{V} T( s\operatorname{Li}_{s+1}(a_0)-\log (a_0) \operatorname{Li}_{s}(a_0)).
\tag{75}
\end{equation}

We note that the dimensionless nonstandard  specific energy $\mathfrak{E}_{sp0}$ depends on
the two variables $\mathfrak{V}$ and $s$,
while the dimensional  nonstandard specific energy ${E}_{sp0}$ depends already
on three variables $\mathfrak{V}$, $s$, and $T$,
where~$T$ is also a dimensional variable.

We have considered  above the behavior of the Bose--Einstein distribution in a neighborhood of the point $a=0$
and showed that the decay of a boson into two fermions occurs at the point $a=a_0$ different from zero.
Then, using an analog of the Gentile statistics for $K=0$,
we calculated the value of the nonstandard specific  energy
required for the transition of a boson into two fermions.
Despite the fact that the Gentile statistics was previously applied to the number of particles greater than~$1$,
the use of the nonstandard analysis (Leibnitz differential or monads)
allowed the author to generalize the Gentile statistics relations
to the case of a small number of bosons for $N=K=0$.

Thus, using mathematical tools, we showed that the application of Gentile statistics to monads
allows one to obtain an approximate answer for the problem of determining the  nonstandard specific  energy
of transition of a boson into two fermions.

The notion of wave packet means that a particle is not a point, but it is spreading.
This process depends on the thermal wavelength of de Broglie wave packets.
If we consider the $\mathfrak{V}$-functions corresponding to nucleons
which are related to the quarks  through the variables
in the symmetry groups with  a large number of degrees of freedom,
then the number of variables can significantly increase.
In this case, one can associate quantum mechanical particles
with monads of nonstandard analysis

\section{Considering the attraction.\\ Dimers (pairs) as observable quantities}

\subsection{Second quantization of classical mechanics and ultrasecond
quantization of thermodynamics. Operators of creation and annihilation for
pairs-dimers}

The second quantization is always associated with the identity of particles,
and, if it is carried out for classical particles, then it is tacitly assumed
that the particles are indistinguishable for the observer. Instead of an
$N$-dimensional problem, we arrive at the three-dimensional picture in which
$N$ particles are distributed. The Vlasov equation~\cite{28, 29}
is obtained from the second quantization of classical mechanics. However, the
original arguments used by Vlasov were actually based on the assumption that
the particles can be regarded as identical ones.

In the classical system of Hamilton equations for $N$ particles, even if the
Hamiltonian is invariant under any permutation of the particles, the initial
conditions need not have this property. However, the initial conditions of the
Liouville equation can be regarded as data satisfying the conditions of
symmetry.

Indeed, let
\begin{equation}
\{q_i^0,p_i^0\}, \qquad (q_i^0=q^0_{i,1},q^0_{i,2},q^0_{i,3},
\qquad p_i^0=p_{i,1}^0,p_{i,2}^0,p_{i,3}^0) \tag{76}
\end{equation}
are initial conditions for the Hamiltonian
system whose Hamiltonian is invariant under every permutation $ p_i, p_j $ and
also under every permutation $ q_i, q_j $. For example, let
\begin{equation}
H(p,q)=\sum\frac{p_i^2}{2m}+\sum\sum V(|q_i-q_j|). \tag{77}
\end{equation}
Substituting the initial conditions into the Hamiltonian~\thetag{77},
we obtain the energy
\begin{equation}
E=H(p^0,q^0).\tag{78}
\end{equation}
The energy is conserved along the trajectories of the Hamiltonian system.

Consider further the Liouville equation corresponding to the Hamiltonian system
\begin{equation}
\frac{\partial\rho}{\partial t}=\{H(p,q),\rho\}, \tag{79}
\end{equation}
where $\{\cdot\} $ stands for the Poisson bracket, with
the initial condition $$ \rho(E), \qquad \rho(E)\in C^\infty, $$
where $E$ satisfies~\thetag{78}. This equation describes the distribution
corresponding to the Hamiltonian system with the initial conditions given
above.

The Liouville equation and the initial conditions are symmetric with respect to
any transposition of $p_i$ and $p_j$ and to any transposition of $q_i$ and
$q_j$. This symmetry is preserved for the solutions. According to the Gibbs
distribution for the Gibbs Ensemble, every distribution can be expressed in
terms of energy. Therefore, it is symmetric with respect to any permutation of
the particles.

Sch\"onberg~\cite{30} carried out a second quantization of this system in the
Fock space\footnote{This space exists only under the assumption that the
particles are indistinguishable (the ``commutativity,'' or the invariance).}.
In~\cite{29}, the Vlasov equation was obtained under the assumption that the
interaction is small and the number of particles is large. The BBKKI chains are
also symmetric with respect to these permutations. Hence, for any distribution
in the many-body problem, such a symmetry follows. Thus, we arrive at the
invariance with respect to the permutations of the particles, and thus to a
``distribution of Bose--Einstein type'' for the statistical physics of
classical particles. Other mathematically rigorous arguments which lead to a
``distribution of Bose--Einstein type,'' in the form of lemmas and theorems
(see~\cite{31}), and hence also the distributions of classical particles, obey
the laws of number theory.

Although a modern macroinstrument cannot trace the motion of every particle
(because of discreteness
of the observation times in relaxationally stepwise process), which is possible at the classical
level, but it can distinguish between molecules and dimers or clusters. The clusters consisting of
more than two molecules occur in gas much less frequently than dimers. Dimers are observable at all
temperatures, and a macroinstrument can calculate their average percentage at a given temperature.
This is an important new phenomenon in experiment, and this phenomenon was not available to the
great who formulated the basic laws of thermodynamics.

Dimers occur and become immediately split by monomers (single molecules), and
they are created and annihilated in different places. They occur because there
are {\it quantum\/} forces of attraction between molecules (the dipole-dipole
interaction). The dimers are virtual, as the ideal liquid is.

To take into account this important phenomenon (the creation and annihilation
of dimers) mathematically, one obviously needs to make the ``second''
quantization and introduce the creation and annihilation operators for dimers,
i.e., for pairs. The author referred to this ``second'' quantization as the
``ultrasecond'' quantization due to the introduction of creation and
annihilation operators for pairs. In the special case of the Bardeen model,
this quantization was introduced in essence~\cite{32}, ~\cite{29};
however,since this model is, roughly speaking, exactly solvable, these
operators turned out to be hidden in the model in a sense.

In general, the ultrasecond quantization and the asymptotic behaviors
associated with it are rather cumbersome and lead to quantum equations
involving the Planck constant $\hbar$. The passage to the limit as $\hbar\to0$,
and then the passage to the limit as the viscosity tends to zero, are
cumbersome, and we present here only a part of this passage, which is related,
as in the previous section, with the introduction of an infinitesimal viscosity
into the classical scattering problem (for $ \hbar = 0 $ ). This means that we
introduce the viscosity and, after manipulations, pass to the limit as the
viscosity tends to zero. This procedure will enable us to find the Boyle
temperature $T_B$, and then also the Boyle density $ \rho_B $, i.e., the
so-called Zeno line, which is present in the van der Waals model and which was
first noticed by Bachinski in experiments with pure gases.

We shall further obtain the so-called law of corresponding states.

\subsection{Boyle temperature as the temperature above which the dimers are
not observable in the Boltzmann--Maxwell ideal gas}

The attraction between the particles occurs in the quantum mechanical
consideration of the dipole-dipole interaction. In the standard semiclassical
limit, if the distance between neutral molecules is fixed (does not depend on
the parameter $\hbar$, i.e., on a dimensionless parameter proportional to
$\hbar$), then, as $\hbar\to0$, the attraction disappears. In this sense, the
use of an attraction potential in molecular dynamics using the classical Newton
equations for many particles is at least baseless.

However, if, along with $\hbar$, the problem involves other small and large
parameters, then the attraction potential can be kept for some relationships
among these parameters under the passage to the limit as $\hbar\to0$.

Since the scattering problem has another parameter tending to infinity, for
example, the time of scattering is considered in the interval from $ -\infty $
to $ +\infty $, it can happen that, as $ \hbar \to 0 $ and $ t \to \infty $
simultaneously (provided that there is a dependence between these parameters),
an attractive potential of the order of $r^{-6}$ is kept (as $\hbar\to0$),
where $r$ stands for the distance between the particles.

As we see below, to obtain a ``rough'' thermodynamics leading
to the law of corresponding states, it is sufficient to determine
the values of the Boyle temperature $T_B$ and the Boyle density $\rho_B$
only for mercury.
As is known, the mercury isotherms are very close to the Van der Waals model
(see Fig.~14),
and hence the Lennard--Jones potential model must provide a good description
of the following two important facts:
1) attraction existence;  2) collision of molecules due to a rapidly increasing
repulsive potential.

We present only a typical example of studying the relation between the actual gas
and the interaction potential, which corresponds to the case of a small intermolecular
distance such that quantum effects must inevitably arise.
But the natural choice of the potential $\Omega$ for an actual gas
and and the fact that the Zeno lines are taken into account
give us mercury isotherms in Sec.~4.5.

As an example, we consider the Lennard--Jones potential, noting that,
in our fundamental
manipulations, the repulsive part of the potential does not play any role.

The only essential quantity is the so-called effective radius $a$, because it determines a
one-dimensional elementary length.

As is known, in the radially symmetric case,
\begin{equation}
\frac{mv^2}2+\frac{M^2}{2mr^2}+\Phi(r)=E. \tag{80}
\end{equation}

In the original scattered particles, we prescribe an energy $E$ and an impact
parameter $B$. The momentum $M$, as well as the energy $E$, is preserved. We
also know that
\begin{equation}
M^2=B^2E. \tag{81}
\end{equation}
Expressing the energy $E$, we obtain for the attraction
\begin{equation}
E=\frac{(mv^2)/2+\Phi(r)}{1-B^2/r^2} \tag{82}
\end{equation}
in the domain $ r \leq B $.

In the scattering problem, for the interaction potential, one considers the
Lennard--Jones potential
\begin{equation}
\Phi(r',r'')=
4\varepsilon\big(\frac{a^{12}}{\|r'-r''\|^{12}}-\frac{a^6}{\|r'-r''\|^6}\big),
\qquad r=r'-r'', \tag{83}
\end{equation}
where $\varepsilon$ stands for the energy at the well depth, $a$ for the
effective radius, and $\|r'-r''\|$ for the distance between two particles with
radius vectors~$r'$ and~$r''$. In the two-particle problem, in the absence of
external potential, the problem is reduced to a one-dimensional radially
symmetric problem.

In problem~\thetag{82}, for different values of $B$, there are other barriers
and wells. At the stationary points $E_{\min}$ and $E_{\max}$, the velocity
vanishes, and thus these values can be evaluated by using the potential term
only.

We speak now of a pair of particles with the mass center which is caught by the
trap (rather than of a single particle). Therefore, the difference
$E_{\max}-E_{\min}$ is the very energy which is needed to knock out the pair
(dimer) from the trap.

After formation of a dimer at $T\leq T_c$, one should consider the
collision of a dimer with a monomer according to the same scheme,
assuming that the pair-dimer (a $\mu$-particle) has already been formed.
Further, one considers the scattering problem already
for the dimer at $E_{\min}=T$ and a particle of the same mass $m$.
The successive consideration of such a multistep procedure
leads to formation of a three-dimensional cluster,
and we obtain a temperature significantly lower than~$T_c$.
It is important that the first step gives an upper bound,
and this restriction is natural.

In an experiment, the percentage of dimers in gas can be calculated. It can be
seen how dimers are created and how they are annihilated (broken by monomers).
After this, the average number of these events is calculated. The higher is the
temperature, the higher is the average energy of the monomers, and the smaller
is the number of dimers.

The main point is that, under this approach, there are only two values,
$E_{\max}$ and $E_{\min}$, which are kept in the skeleton of the scattering
problem (cf.~the skeleton of the amoeba in the tropical mathematics~\cite{24}).
For $E_{\max}=E_{\min}$, the well disappears. For the attractive part of the
Lennard--Jones potential, this energy is equal to $0.8\varepsilon$. With regard
to the standard Clausius considerations\footnote{Following Clausius, experts in
molecular physics usually argue by proceeding from the symmetry of the motion
of a molecule in all six directions. In the scattering problem, we use
the principle of symmetry in all directions, which is standard in molecular
physics. The fraction of all particles that moves head-on is 1/12.
There are three such directions; hence, one quarter of all
molecules collide. The arguments concerning symmetry that were used by Clausius
to evaluate the free path length (and are repeated here by the author) are
quite approximate. However, these arguments do not modify the values of the
ratios of the form $T_B/T_{c}$. This very ratio is of interest for us.},
we can see that the average energy of the particles is equal to
$\frac{16}{5}\varepsilon$. The average energy is the temperature, and
$T=\frac{16}{5}\frac{\varepsilon}{k}$. Above this temperature, there is no
well. In thermodynamics, for physical reasons, this is the so-called Boyle
temperature $T_B$. In our framework, the Boyle temperature is defined as the
temperature above which the dimers are practically absent. This is a new
approach. According to this conception, the Boyle temperature for argon (Ar) is
$T_B=382\,K$ and for krypton (Kr) it is $T_B=547\,K$, while the tables of the
experimental work~\cite{33} give $T_B=392\,K$ for Ar and $T_B=538\,K$ for Kr.
The discrepancy between the theoretical and experimental values is of the order
of 2--3\%.

The critical temperature $E_{\max}$ must correspond to the deepest well, i.e.,
to the maximum value of the difference $E_{\max} - E_{\min}$ for all impact
parameters $B$. This difference determines the drop of the energy of a dimer
after this dimer was captured by the ``trap,'' and thus determines the energy
which a monomer must have to knock out the dimer from the well (i.e., for the
dimer to collapse).

The height of the barrier ``protects'' the created pair whose reduced mass was
captured by the trap of ``dimers'' and ``clusters''  from ``shocks'' of
monomers. As the temperature decreases, $ T <T_{c} $, the height of the barrier
reduces, and, to survive, the clusters must create their own barrier in the
form of a microanalog of the surface film. Thus, a ``domain'' must occur, a
three-dimensional cluster (the so-called elementary cluster) which has at least
one particle which is protected by other particles.

This is a new definition of the critical temperature $T_c$ as the temperature
below which clusters are formed from dimers. Calculations give $E_{\ max}=0.286
\frac{\varepsilon}{k}$ at the point $\max_B(E_{\max}-E_{\min}) $. The impact
parameter at this point is equal to $B=2.436 a$.

\subsection{Macroinstruments and microinstruments in dimension theory.
determinin the maximum density $\rho_B$ and the Zeno line which borders the
domain of dimers as the density is modified}

Let us now obtain analytical formulas for the Zeno line in dependence on the
potential.

\renewcommand{\thefigure}{\arabic{figure}}

\begin{figure}[h!]
\begin{center}
\includegraphics{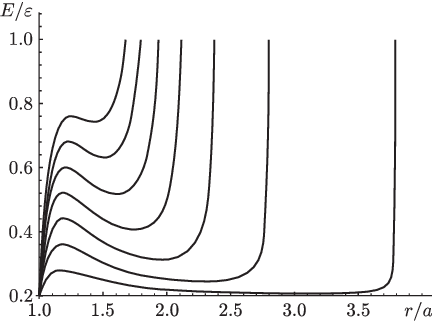}
\end{center}
\caption{Wells and barriers in the scattering of two particles with the Lennard--Jones
interaction potential at different impact parameters~$B$.}
\end{figure}

\begin{figure}[h!]
\begin{center}
\includegraphics{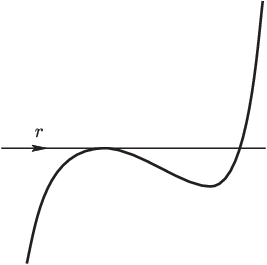}
\end{center}
\caption{A trap for a fictitious particle--dimer, in the center of mass system (CMS).
Here $r$ stands for the radius vector of the dimer; it is marked on the abscissa
axis. A ``particle'' falls from the left, from the point $r=B$, where $B$
stands for the impact parameter. After reflection from the ``wall'', i.e., from the potential,
in the presence of viscosity, the particles meet the barrier and
then after rather many reflections from the barrier and the ``wall''
drop on the well bottom in a greater time interval.}
\end{figure}

Let us use considerations of dimensional analysis for the scattering problem
and the definition of the one-particle (the so-called thermal) attraction
potential. The scattering problem is considered for the Lennard--Jones
potential, and therefore there is an additional parameter of length in the
problem, the parameter $a$, which is the effective radius. The attraction
potential occurs in the quantum theory of the dipole--dipole interaction. If we
fix the distance between the particles and assume that the semiclassical
parameter $\hbar$ tends to zero, then the attractive potential vanishes. This
means that, as $\hbar\to0$, the distance between the particles decreases. It
follows that the attractive potential acts between ``nearest neighbors'' only.

Therefore, it is natural to use the expansion of an attraction single-particle
potential in powers of the radius $r$ up to $O(r^3/V)$ only.

The ``dressed'' or thermal potential $ \Psi(r^2) $ is attracting. As is well
known, it was derived from the fact that the correlation sphere for the
$N$-particle for the Gibbs distribution is finite~\cite{34}.

One of the most interesting points of independence of a macroinstrument of a
microinstrument manifests itself when applying the dimension theory~\cite{35}.
A macroinstrument determines the volume $V$. According to the independence of
the thermodynamic quantities on the shape of the volume~$V$, the volume $V$
ensures us that we have the dimension in the dimension theory is three;
however, this volume does not give us any one-dimensional measure. which is the
typical length in the thermodynamic process.

A microinstrument determines the effective radius of the molecule and the mean
free path. However, in dimension theory, we cannot measure the typical length
of the macrothermodynamics by using the radius of the molecule or the mean free
path, even if the volumeis a Torricelli tube and its typical thickness is small
as compared with the case in which the vessel is a ball. These considerations
show that the only possible dimensionless combination for the argument of
$\Psi(r^2) $ is
$$ \Psi\Big(\frac{ar^2}V\Big).
$$ Since $ a^3 \ll V $, it follows that the
expansion is
\begin{equation}
\Psi\Big(\frac{ar^2}V\Big)=C_1+
C_2\frac{ar^2}V+O\bigg(\frac{a^2r^4}{V^2}\bigg). \tag{84}
\end{equation}
The constant $C_1$ gives no contribution to the scattering problem, and the
thermal ``single-particle'' potential turns out to be proportional to the
density. On the plane $\{T,\rho\}$, the maximum of the binodal (according to
Fig.~8 for $T=T_c$) is equal with respect to $\rho$ to the very value $\rho_c$,
which enables us to find the proportionality coefficient. It turns out to be
equal to one.

The situation in which the thickness of the tube is ``commensurable'' to the
radius of the molecule\footnote{When the scales become ``commensurable'' in
this sense, another thermodynamics arises~\cite{27, 36}.} leads to quite
different effects: to the superfluidity of water in a nanotube and to the
freezing of water at $T^0 = 5K$ (see~\cite{36}).

\begin{figure}[h!]
\begin{center}
\includegraphics{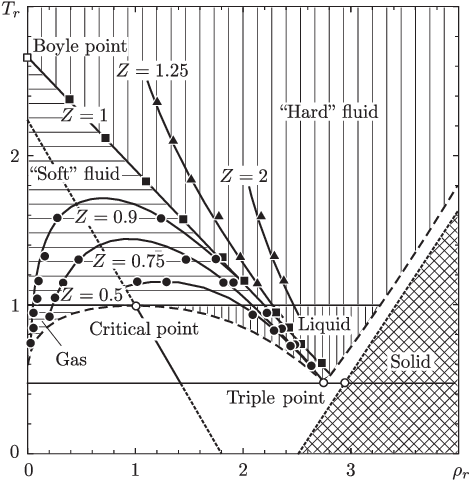}
\end{center}
\caption{The $T$--$\rho$-diagram for gases corresponding to simple liquids. Here
$T_{r}=T/T_{c}$ and $\rho_{r}=\rho/\rho_{c}$. The Zeno line (the straight line
$Z=\frac{PV}{kNT} = 1.0$) in the phase diagram. For the states $ Z>1.0 $ (the
hard fluids), the repulsive forces dominate; for the states $ Z <1.0 $ (soft
fluids), the attraction forces dominate.}
\end{figure}

Expanding
\begin{equation}
r^2=r^2_1+r^2_2=\frac{(r_1-r_2)^2}2+\frac{(r_1+r_2)^2}2, \tag{85}
\end{equation}
we can make the separation of variables in the two-particle problem to a
scattering problem for a pair of particles and to the problem of their joint
motion for $r_1 + r_2$, just as it was done in \cite{37}. In this case, in the
scattering problem, a quadratic attraction potential (an upturned parabola)
$-\rho r^2$ is added to the Lennard--Jones interaction potential, and $
\rho=\frac 1V$ for the isochoric process.

In the scattering problem thus obtained, there are two points of rest, namely,
the stable one, $E_{\min}$, and the unstable point $E_{\max}$. Their ratio is a
dimensionless quantity. As follows from the previous section,
$Z=\frac{PV}{NT}$, where $P$ stands for the pressure, $N$ for the number of
particles, $T$ for the temperature, and, due to the fact that a stable
stationary point has the meaning of temperature, it follows that the ratio
\begin{equation}
Z=\frac{PV}{NT}=\frac{E_{\min}}{E_{\max}} \tag{86}
\end{equation}
enables us to wrote the curves $ Z = \text{const} $ in the graph $T$,
$\rho=\frac{N}{V} $.

The curve at $Z=1$ is called the {\it Zeno line} (or the {\it Bachinski
parabola}), and the locus of the beginnings of the curves $Z_{\max}$ (for
$C_2\neq 0$ and $B\to\infty$) is referred to as the {\it binodal}.

Thus, in our view, the Zeno line determines the temperatures for which the
dimers become practically nonexistent for a given density.

Denote by $\rho_{c}$ the value of $\rho$ at the maximal point of the binodal
and denote the endpoint of this curve on the $\rho$ axis by $\rho_B$.
In~\cite{38}. this point was referred to as the hypothetical point $\rho_B$
(the Boyle point).

Calculating the value of $Z_{c}$, we obtain $Z_{c} = 0.296$, which coincides
with the values of $Z_{c}$ for the noble gases up to thousandths. The ratio
$\rho_{c}/\rho_B$ also coincides with the values of this quantity for the noble
gases.

Table~1 shows the data corresponding to the resulting diagram (for $B=100$ in
``molecular'' values), and note the discrepancy between the basic dimensionless
relations obtained by the data of molecular dynamics and the theoretical
relations obtained by physicists from the chain BBKKI and the $N$-partial Gibbs
distribution.

\bigskip
Table 1

$$
\begin{matrix} & Z_{c}\quad &\rho_{c}/ \rho_B \quad &T_{c} / T_B \\
&0.29 \quad &0.273 \quad &0.36\\
&0.308 \quad &0.285 \quad &0.38\\
&0.375 \quad &0.333 \quad &0.296
\end{matrix}
$$

On the top line of the table, the theoretical values for $Z_{c}$,
$\rho_{c}/\rho_B$, and $T_{c}/T_B $, obtained using the above theory are
presented. The second line contains the values of the same quantities evaluated
according to the latest data of molecular dynamics and results of theoretical
physicists for the Lennard--Jones potential. The third line gives the values
obtained from the van der Waals equation, which is empirical.

The value of $ Z_{c} $ can be computed in the experiment very accurately, and
it is equal to $0.29$ for noble gases, nitrogen, oxygen, and propane. The value
of $\rho_{c}/\rho_B$ (the ratio of the critical value of $\rho$ critical to
$\rho_B$, i.e., to the entire length of the segment with respect to $ \rho $ on
which the Zeno-line ``cuts'' the abscissa axis away) evaluated in the above
theory coincides with the corresponding values for water, argon, xenon,
krypton, ethylene, and a number of other gases.

 Let us present detailed calculations to find the Zeno-line.

Consider the potential
\begin{equation}
E(r)=\frac{-\alpha r^4+r^2U(r)}{B^2-r^2}. \tag{87}
\end{equation}
Its first derivative is equal to
\begin{equation}
E'(r)=\frac{r(2B^2U(r)+r(2\alpha r(-2B^2+r^2)+(B^2-r^2)U'(r)))}{(B^2-r^2)^2},
\tag{88}
\end{equation}
and the second derivative is
\begin{align}
E''(R)&=\frac1{(B^2-r^2)^3}\bigl(2(B^4+3B^2r^2)U(r)+r(-2\alpha r(6B^4-3B^2r^2+r^4)
\notag\\
&\qquad +4(B^4-B^2r^2)U'(r)+r^2(B^2-r^2)^2U''(r))\bigr).
\tag{89}
\end{align}

We obtain a solution of the equation in the form
\begin{equation}
B=\sqrt{-\frac{-r^3U'(r)+r^4U''(r)}{-8U(r)+2rU'(r)+2r^2U''(r)}}. \tag{90}
\end{equation}

Substituting the value $B(r)$ into~\thetag{87}, we find $E(\alpha)$, the
Zeno-line, i.e., a segment, which  is straight up to 3\%,
$T/{T_B}+\rho/{\rho_B}=1,$ where $\rho_B$ stands for the maximal density as
$T\to0$.

\setcounter{figure}{7}
\renewcommand{\thefigure}{\arabic{figure}A}

\begin{figure}
\begin{center}
\includegraphics{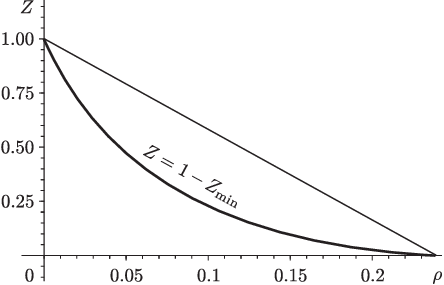}\\
\end{center}
\caption{Curve $1-Z_{\min}$.}
\label{fig_8a}
\end{figure}

\subsection{Limit stretching of a liquid.\\ The maximum density of holes}

We have $ Z= P/{\rho T}, $ where $ \rho=\rho_B\left(1-
T/{T_B}\right)$ is an isotherm-isochore. Therefore,
\begin{equation}
Z=\frac{P_r}{\rho_BT_r(1-T/T_B)}. \tag{91}
\end{equation}
Let us express $Z$ in terms of $\gamma $ for $Z <0$ and $\gamma<0$ and for
$\mu\sim o (1/\ln N)$,
$$
N=A(\gamma)T_r, \qquad
P=\frac{T_r^{2+\gamma}\zeta(2+\gamma)}{\zeta(2+\gamma_c)},
$$
where $T_r=T/T_c$ and $P_r=P/P_c$. The
value of $\rho_B $ is defined in this very normalization.

Therefore,
\begin{equation}
Z=\frac{T_r^{2+\gamma}\zeta(2+\gamma)}{\zeta(2+\gamma_c)(A(\gamma)T^2}=
\frac{T_r^\gamma\zeta(2+\gamma)}{\zeta(2+\gamma_c)A(\gamma)}. \tag{92}
\end{equation}
On the other hand, $ N=\rho=\rho_B\left(1-T/{T_B}\right)=A(\gamma)T. $
Consequently,
${\rho_B}/T-{\rho_B}/{T_B}=A(\gamma)$ and
$$
\left(\frac1T\right)^{|\gamma|}=
\left(\frac1{\rho_B}A(\gamma)+\frac1{T_B}\right)^{|\gamma|}.
$$
Substituting this into~\thetag{92}, we obtain
$$
Z=\frac{\zeta(\gamma+2)}{\zeta(\gamma_c+2)}
\frac{\left(\dfrac1{\rho_B}A(\gamma)+\dfrac1{T_B}\right)^{|\gamma|}}{A(\gamma)},
$$
and $ A(\gamma) \to \infty $  as $ \gamma \to -1 $.

Therefore, as $ \gamma \to -1 $, the value $ 1/T_B $ is negligible.
Consequently,
\begin{equation}
Z|_{\gamma\to-1}\simeq
\frac{\zeta(\gamma+2)}{\rho_B^{|\gamma|}\zeta(\gamma_c+2)A(\gamma)^{1-|\gamma|}}.
\tag{93}
\end{equation}
Here $\gamma_c$ corresponds to $Z_c$, which is the minimal value of $Z$ on the
critical isotherm for $P=1$ (see Fig,~9).

Since
$$
A(\gamma,\Lambda)=(\Lambda^{\gamma_c-\gamma}c(\gamma))^{1/1+\gamma},
$$
where $ \gamma <0 $, it follows that
$$
A(\gamma,\Lambda)^{1+\gamma}=\Lambda^{\gamma_c-\gamma}c(\gamma),
$$
where
$$
c(\gamma)=\left[\int_0^\infty
t^\gamma\,dt\left(\frac1t-\frac1{e^t-1}\right)\right].
$$
Hence,
$$
Z|_{\gamma\to-1}=
\frac{\zeta(2+\gamma)}{\rho_B^{|\gamma|}\zeta(2+\gamma_c)c(\gamma)\Lambda^{\gamma_c+1}}.
$$

The expression $c(\gamma)$ tends as $\gamma\to -1 $ to $(1/2)\ln\varepsilon$,
where $\varepsilon$ stands for the lower limit of the integral expression for
$A(\gamma)^{1-|\gamma|}$ at $ \gamma = -1 $.

Similarly, the expression $\zeta(\gamma +2)$ at $\gamma=-1$ is equal to
$\ln\varepsilon$.

Hence
$$
Z|_{\gamma\to-1}=\frac2{\rho_B\zeta(\gamma_c+2)\Lambda^{\gamma_c+1}}
$$
as $\gamma\to0$ and $Z\to0$. Therefore,
$$
Z<Z_{\max}< \text{const}
$$
for all values of $\gamma$.

Hence, by~\thetag{91},
\begin{equation}
P_r=Z\rho_BT_r\left(1-\frac{T_r}{T_B}\right)\le Z_{\max}\frac{T_B}2. \tag{94}
\end{equation}
Moreover, it is clear that $P_r\to0$ as $T_r\to0$ and
$\gamma\to0$.

The value of $ \lambda $ determines the minimum of $P_r$
and the maximal density of holes.

To determine the gas-liquid transition, as in Sec.~2.3,
with the correction to the Zeno line taken into account,
we derive equations of the form~\thetag{25}-\thetag{26},
by normalized the activity for the critical isotherm.

\renewcommand{\thefigure}{\arabic{figure}}

\begin{figure}[h!]
\begin{center}
\includegraphics{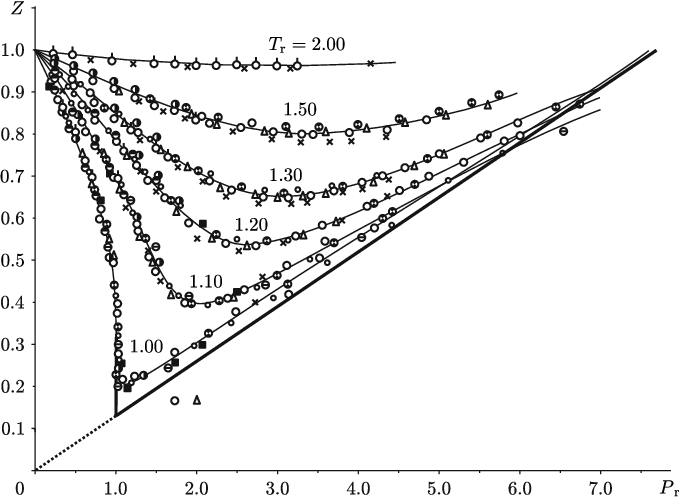}
\end{center}
\caption{The picture of the law of corresponding states for molecules of different
gases. Thin lines show isotherms for methane. Different symbols on isotherms
correspond to argon, carbon dioxide, water, etc. The fact that isotherms of
different gases are close to one another illustrates the empirical law of
corresponding states. The theoretical isotherm (the solid line) does not fully
coincide with the experimental one. This is an effect of the same type as the
jump of the critical exponents. The viscosity (the Wiener quantization) smooths
the sharp angle of the limit isotherm (as $\varepsilon\to 0 $).}
\end{figure}

\begin{figure}[h!]
\begin{center}
\includegraphics{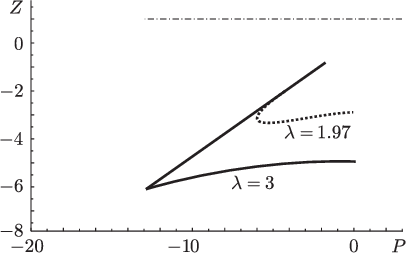}
\end{center}
\caption{Spinodal curves in the domain $P<0$. The inclined line is the continuation of
the theoretical critical isotherm shown in Fig.~9 to the domain of negative
values of $P$ and $Z$. The initial point of the curve $ \Lambda = 1/3$
($\lambda=1/\Lambda$) is at a distance from the point $P=0$, and the starting
point $\Lambda=0.5086$ ($\lambda=1/\Lambda=1.97$) of the curve coincides with
the point of intersection of the dashed line and the solid line.}
\end{figure}

\subsection{A coarser measuring instrument\\ and the law of corresponding states}

Obviously, the rougher is the device, the greater is the dissipation parameter,
and therefore the more important is the uncertainty principle. If our device
does not distinguish some molecules from one another, then this is a rough
instrument. It is not necessary that the device itself be so rough. It is
sufficient to say that the experimenter does not want to distinguish molecules
and computes the density roughly, counting all the molecules for which he wants
to construct a ``rough'' thermodynamics. Hence, when speaking more precisely,
from the point of view of mathematics, the rougher thermodynamics we want to
receive, the more rough will be the answer due to the Wiener uncertainty
principle.

Our rough instrument does not distinguish particles not only in mixtures.
Measuring different particles, the observer thinks that these are the same
molecules, and only the measurements are somewhat scattered. This is even a
more rough thermodynamics. It is referred to the fluid part, when the
gas-liquid is not distinguished, clusters occur, etc. The critical points are
on the boundary of the fluid domain, and we approach them from the side of the
fluid domain for $ T \ge T_c $.

When considering above the phase transition, we have equated the chemical
potentials of the liquid and the gas phases on the critical isotherm, assuming
that there is no phase transition there. We considered the case in which the
viscosity vanishes, and we obtained a phase transition which was not smeared.
Now, when considering the problem with a coarser device, we no longer have any
right to assume that the dissipation tends to zero. The uncertainty principle
gives us a fairly large smearing of the phase transition; however, it happens
on a ``rough'' critical isotherm which is measured by a rough instrument. The
latter cannot distinguish now not only particles of a single pure gas but does
not distinguish particles of different gases either.

We have compared the Wiener quantization of thermodynamics with the dissipation
resulting in a shock wave. However, the dissipation may be different in
different substances, while we are interested now in the Hugoniot conditions
for the entire mixture, and we do not want to distinguish stratificationally
occurring internal shock waves within a common shock wave. This is particularly
evident when the dissipative parameters of viscosity and thermal conductivity
are different (see~\cite{40}, \S~95), the viscosity $\nu$ is small, and the
thermal conductivity is relatively large, $\chi\gg\nu$. If the processes inside
the shock wave are not of interest for us, then we pose the Hugoniot conditions
on a shock wave spread with respect to heat conductivity.

On the other hand, if the values of viscosity in a mixture of different gases
are different but not dramatically different from one another, then the width
of a shock wave thus obtained is defined by the average viscosity.

Since, as a rule, the Bose--Einstein-type distribution is considered in the
three-dimensional case, it would seem to be natural to add the integration over
the coordinates to the integration of the momenta. Then the three-dimensional
volume $V$ would occur as if this is a natural way. It would seem that there
can be a generalization if the dimension of volume is changed when changing the
dimension of the momenta. However, in the manual~\cite{2}, when considering a
photon gas, the integration of the Hamiltonian of the oscillator is carried out
both over the momenta and over the coordinates, and, nevertheless, the
three-dimensional volume $V$ is taken as the multiplier for the distribution.

Note first that number theory gives, for dimension two, a distribution without
the volume $V$ (see Example~1), as well as the initial relations in~\cite{2}
both for the Boltzmann--Maxwell ideal gas and for the Bose--Einstein ideal gas
(see \thetag{1}--\thetag{4}). Further, in~\cite{2}, both the distribution for
bosons and the distribution for photons are multiplied by the three-dimensional
volume $V$. Certainly, the main distribution is the distribution without the
volume $V$, and its multiplication by $V$ is caused by the correspondence with
thermodynamics in which the pair `the volume $V$ -- the pressure $P$' is the
most important tool.

Therefore, the most natural generalization to the nonideal distribution is the
multiplication of a fractional Bose distribution by a function of $V$ of the
form $V\varphi(V/V_0)$, where $V_0$ stands for some reference volume and the
function $\varphi(x)$ is smooth.

The introduction of this multiplier does not change the distribution caused by
number theory~\cite{41} in which the variable $V$ is eliminated by the change
of the variable $N/V=\rho$. However, if we consider $V\varphi(V/V_0)$  as a
multiplier, where $V_0$ is some typical volume, then the relation for the
three-dimensional Lagrangian manifold $\Lambda^3 $ in the six-dimensional phase
space $\{P,V;T,S;\mu,N\}$ is preserved, whereas the variables $V$ and $N$ do
not convert here into a single variable $\rho=N/V$. Therefore, the volume and
the number of particles are changed on the isochore $\rho=\text{const}$ in
general. This modification of the $\Omega$-potential does not change the
specific entropy, which is also of importance.

In a mixture of gases, we are to choose a reference gas in which the difference
between the vapor and the liquid is the lowest possible, for example, from the
point of view of the number of dimers. This gas is the mercury vapor
($Z_c=0.4$). Let us carry out a normalization of activity~\thetag{26} for an
isotherm of this gas at $Z=0.4$ and assume that there is no phase transition on
the critical isotherm of mercury. In accordance with~what was said above, the
rough device cannot distinguish among molecules of $ l $ distinct gases.
Let us calculate the average degree of freedom for these molecules by taking
the arithmetic mean of the values of the entropy of $l$ pure gases
(see~\thetag{13}) on the basis of experimental data for
$Z_c^i$, $i=1$, $2$, \dots, $l$, pure gases
(see~\thetag{13}),\vskip-1pt\noindent
$$
(\gamma_{average}+2)
\frac{\zeta(\gamma_{average}+2)}{\zeta(\gamma_{average}+1)}=
Z_{average}(\gamma_{average}+2)=\frac1l\sum_{i=1}^l(\gamma_i+2)
\frac{\zeta(\gamma_i+2)}{\zeta(\gamma_i+1)}.
$$\vskip-4pt\noindent
The highest value $Z_c=0.4$ is given by mercury (Hg), and therefore the average
number of degrees of freedom of this family of molecules is certainly less, and
therefore $Z_{average}<Z_c$ (mercury), $\gamma<\gamma_{Hg}$. The critical
pressure is greater than that for mercury,
$\zeta(\gamma_{Hg}+2)<\zeta(\gamma_{average}+2)$.. Therefore, the value we have
chosen for mercury, $P_r=1$, is less than
$P_{average}=\zeta(\gamma_{average}+2)/\zeta(\gamma_{Hg}+2)$. Thus, the value
$P_r=1$ for $Z_{Hg}$ belongs to the domain of the phase transition
``gas-liquid'' for $Z_{average}$. This implies that, for $P_r=1$, the phase
transition to liquid occurs at $ Z_{average} $.

This phase transition to liquid can be seen in Fig.~9, in the form a vertical
bounded by a black sloping line depicting the liquid\footnote{One can
rigorously prove the existence of phase transition only for the transition of a
new ideal gas into a new ideal liquid without taking into account the Zeno
line, which is unknown for $\gamma_{average}$. Therefore, a rigorously proven
transition from $Z=0.4$, $P=1$ to a liquid is obtained a bit higher than at the
point $Z=0.12$, $P=1$ in Fig.~10.}.

Taking into account the Zeno line influences the form of the $\Omega$-potential
as follows:\vskip-1pt\noindent
\begin{equation}
\Omega(\mu,T)=
-\Lambda^{\gamma+1}V\varphi(V/V_0)\frac{T^{\gamma+2}}{\Gamma(\gamma+2)}
\int_0^\infty\frac{t^{\gamma+1}\,dt}{(e^t/y)-1}=
-\Lambda^{\gamma+1}T^{\gamma+2}V\varphi(V/V_0)\operatorname{Li}_{\gamma+2}(y),
\tag{95}
\end{equation}
where $y=\exp(\mu/T)$ is the activity and $\mu$ stands for the chemical
potential.

Let us write out the differential equations for $\varphi(x)$ with regard to the
relations on the Zeno line,
\begin{equation}
T_z=T_B\left(1-{\rho_z}/{\rho_B}\right), \qquad
P_z=\rho_zT_B\left(1-{\rho_z}/{\rho_B}\right), \tag{96}
\end{equation}
where the subscript $z$ means that the corresponding values are taken on the
Zeno line, i.e., for $ Z = 1 $.

Let us construct the relation $Z=1$ on the Zeno line. This relation is of the
form
\begin{equation}
Z=\frac{\partial\Omega/\partial V}{T_z\partial\Omega/\partial\mu}=
\frac{\varphi(V_z/V_0)+(V_z/V_0)\varphi'(V_z/V_0)}{\varphi(V_z/V_0)}\cdot
\frac{\operatorname{Li}_{\gamma+2}(y_z)}{\operatorname{Li}_{\gamma+1}(y_z)}=1.
\tag{97}
\end{equation}
It follows from~\thetag{96} on the Zeno line that
\begin{equation}
T_z^{2+\gamma}[\varphi(V_z/V_0)+(V_z/V_0)\varphi'(V_z/V_0)]
\operatorname{Li}_{\gamma+2}(y_z)=
\left(N/{V_z}\right)T_B\left(1-N/({V_z\rho_B})\right). \tag{98}
\end{equation}
Assume that the conditions $N/V=\text{const}$ and $N=\text{const}$ hold on the
isochore and on the Zeno line defined by
relation~\thetag{96}.
It follows from~\thetag{97} and \thetag{98} that
\begin{equation}
T_B^{\gamma+1}\left(1-N/({V_z\rho_B})\right)^{\gamma+1}
\varphi(V_z/V_0)\operatorname{Li}_{\gamma+1}(y_z)=N/({V_z\rho_B}). \tag{99}
\end{equation}
Here $N/V_z=\text{const}$ and $N=\text{const}$, and hence this equation
contains unknowns $y_z$, $V_z$, and $\varphi(V_z/V_0)$.

After finding the value $ y_z $ as a function of $V_z$ and $\varphi(V_z/V_0)$,
we substitute it into formula~\thetag{97}, thus transforming
equation~\thetag{97} into a differential equation for the function $\varphi(x)$
depending on the constants $\rho_B$ and $T_B$.
The equation for $\varphi(V)$ enables one to find the point
$y_z(\rho)$ on the Zeno line, and, after this, the isochore is changed only at
the expense of the modification of activity $a=e^{-\mu/T}$ from $y_z(\rho)$ to
$a=1$ for the polylogarithm $\operatorname{Li}_{\gamma+1}(a)$. The function
$\varphi(V)$ is not reflected in the structure of the isochore, and it is
reflected in scaling only (see Fig.~11).

The most important problem in the theory of differential equations, the
existence problem for a solution, remains open. Physicists pay less attention
to this problem than mathematicians.

What are conditions for the existence of a solution of
equations~\thetag{97}--\thetag{98}?

Introduce the notation $T_s=T_{ \text{standard}}=T/T_m$ and
$P_s=P_{\text{standard}}=P/P_m$, where $T_m$ and $P_m$ are defined by the
formulas $$P_m=T_m^{\gamma+2}, \qquad V_0=V_m=V_{\max},$$ and $Z_m=Z_{\max}$
and $\rho_m=\rho_{\min}$ are defined below.

Since $N_c$ does not depend on $T$, it follows that $V$ and $N$ are constant
along the isochore $V=V_m$.

Let us write out the above relations at the point $ T_s = 1 $, $ P_s = 1 $:
\begin{equation}
[\varphi(V_z/V_0)+(V_z/V_0)\varphi'(V_z/V_0)]\zeta(\gamma+2)=1, \qquad
V_m\varphi(1)\zeta(\gamma+1)=N_c, \qquad V_m=V_{\max}, \tag{100}
\end{equation}
Hence,
\begin{equation}
Z_m=\frac{V_m}{N_c}\frac{\zeta(\gamma+1)}{\zeta(\gamma+2)}\cdot
\frac{\operatorname{Li}_{\gamma+2}(1)}{\operatorname{Li}_{\gamma+1}(1)}=
\frac{V_m}{N_c}=\frac1{\rho_m}. \tag{101}
\end{equation}
Since we construct isochores $V=\text{const}$ on the plane $\{Z, P\}$, it
follows that $V_m=V_z$. Eliminating $(V_z\varphi(V_z/V_0)'$ from~\thetag{100} by
using~\thetag{99}, we obtain
\begin{equation}
T_B^{\gamma+1}\left(1-\frac1{Z_m\rho_B}\right)^{\gamma+1}
\frac1{\zeta(\gamma+2)}\operatorname{Li}_{\gamma+2}(y_z)=\frac1{Z_m}. \tag{102}
\end{equation}
Since $V_m=V_z$, we see that
\begin{equation}
\varphi(1)=N_c(\zeta(\gamma+1)V_m)^{-1}, \qquad
\varphi(1)+\varphi'(1)=(\zeta(\gamma+2))^{-1}, \tag{103}
\end{equation}
and it follows from~\thetag{97} on the Zeno line that the following equation
holds:
\begin{equation}
\frac{\operatorname{Li}_{\gamma+2}(y_z)}{\operatorname{Li}_{\gamma+1}(y_z)}
\cdot\frac{V_m}{N_c}\cdot\frac{\zeta(\gamma+1)}{\zeta(\gamma+2)}=1. \tag{104}
\end{equation}
Eliminating $y_z$, we find a relation for $Z_m$ and $\gamma$. The maximum value
of $Z_m$ depends on the values of $\rho_B$ and $T_B$ only. For mercury, this
maximum is obtained for $\gamma_m=\gamma_{\min}=0.1$, and $Z_m=0.4$, which
coincides with the value of $Z_c$ for mercury. This coincidence, which depends
on $\rho_B$ and $T_B$, holds for mercury Hg only (among all the elements of the
periodic table), which confirms the correct choice of the
$\Omega$-potential~\thetag{95}.

The family of isochores, according to system~\thetag{98}--\thetag{99} with the
above initial condition~\thetag{103}, is shown in Fig.~11.

The first relation for the limit isochore $V_m/N_c=Z_{m}$, for $\gamma
=\gamma_{m}$, is of the form
\begin{equation}
Z=Z_m\cdot\frac{\zeta(\gamma+1)}{\zeta(\gamma+2)}\cdot
\frac{\operatorname{Li}_{\gamma+2}(y)}{\operatorname{Li}_{\gamma+1}(y)}, \qquad
1\le y\le y_z(\gamma). \tag{105}
\end{equation}
Since $P_s=T_s^{\gamma+2}\operatorname{Li}_{\gamma+2}(y)/\zeta(\gamma+2)$ and
$N=T_s^{\gamma+1}\operatorname{Li}_{\gamma+1}(y)V_m$, it follows that the other
relation is
\begin{equation}
P_s=\left(\frac{\zeta(\gamma+1)}{
\operatorname{Li}_{\gamma+1}(y)}\right)^{(\gamma+2)/(\gamma+1)}
\frac{\operatorname{Li}_{\gamma+2}(y)}{\zeta(\gamma+2)}, \qquad 1\le y\le
y_z(\gamma). \tag{106}
\end{equation}
Equations~\thetag{105} and \thetag{106} give an almost straight segment of the
isochore.

Starting from $Z<0.4$ (for example, for a van der Waals gas), the phase
transition to a liquid occurs for indistinguishable particles of the law of
corresponding states. This gives a broad area (at the expense of the
uncertainty principle of a ``rough device'') around the line segment $P=1$,
$T_s=1$, $Z<0.4$.
\medskip

{\bf Remark~5.} Since the rightmost isochore (which is not shown in Fig.~11)
is a segment of a straight line, it follows that all isochores of high density
must also be line segments. They pass through the point $\rho>\rho_m$ on the
Zeno line and the point $Z_0=1/\rho$ on the line $P_s=1$. We thus obtain (when
including the isochores shown in Fig.~11) a complete family of isochores for
$Z\leq1$, $P\ge1$, related to the law of corresponding states. To any point of
an isochore in the plane $\{Z, P \}$ there corresponds a point of temperature,
and we construct isotherms which, up to the Wiener uncertainty principle,
correspond to the experimental law of corresponding states (Fig.~12).
\medskip

In Fig.~13, the graph of an experimental isotherm for mercury is presented;
this graph was kindly evaluated by Professor V.~S.~Vorob'ev, according to the
most recent data, at the instance of me. Note that the passage gas--liquid
happens at $T=1473\, K$ along a slanting line rather than a vertical one, which
is related to nonzero viscosity and the Wiener uncertainty principle. This
effect is of the same nature as the jump of critical exponents and ``thickness
of the layer'' of a shock wave.

\begin{figure}[h!]
\begin{center}
\includegraphics{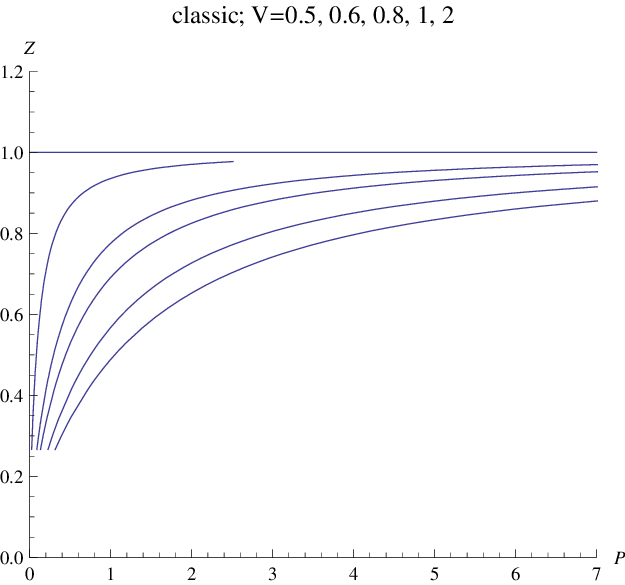}
\end{center}
\caption{The graph for the isochore of the polylogarithm
$\operatorname{Li}_{\gamma+1}(a)$ for $\gamma=2$.}
\end{figure}

\begin{figure}[h!]
\begin{center}
\includegraphics{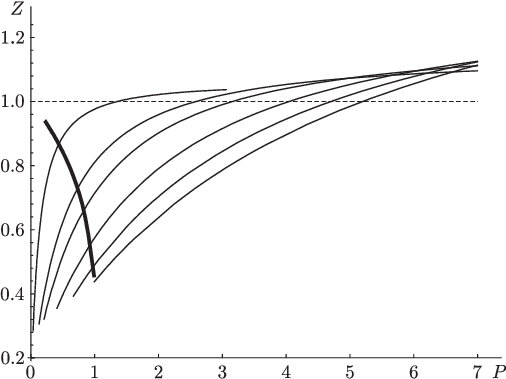}
\end{center}
\caption{The dotted line shows the
Zeno line $ Z = 1 $. The bold line is the critical isotherm of a real gas
(mercury) which is calculated theoretically, and the fine lines are isochores
of mercury for $ T <T_{\text{c}} $.}
\end{figure}

\begin{figure}[h!]
\begin{center}
\includegraphics[width=10cm]{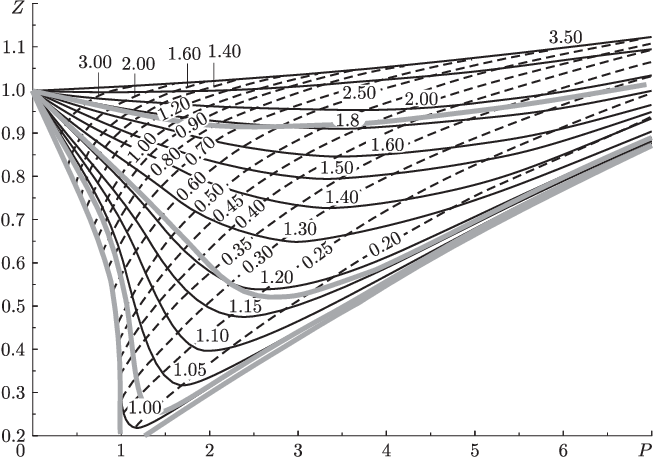}
\end{center}
\caption{The
thin solid lines represent the experimental isotherms for methane (see Fig.~9).
The dotted lines approximating the experimental curves are based on theoretical
data. The dashed curves show the experimental isochores.}
\end{figure}

\begin{figure}[ht!]
\begin{center}
\includegraphics[width=10cm]{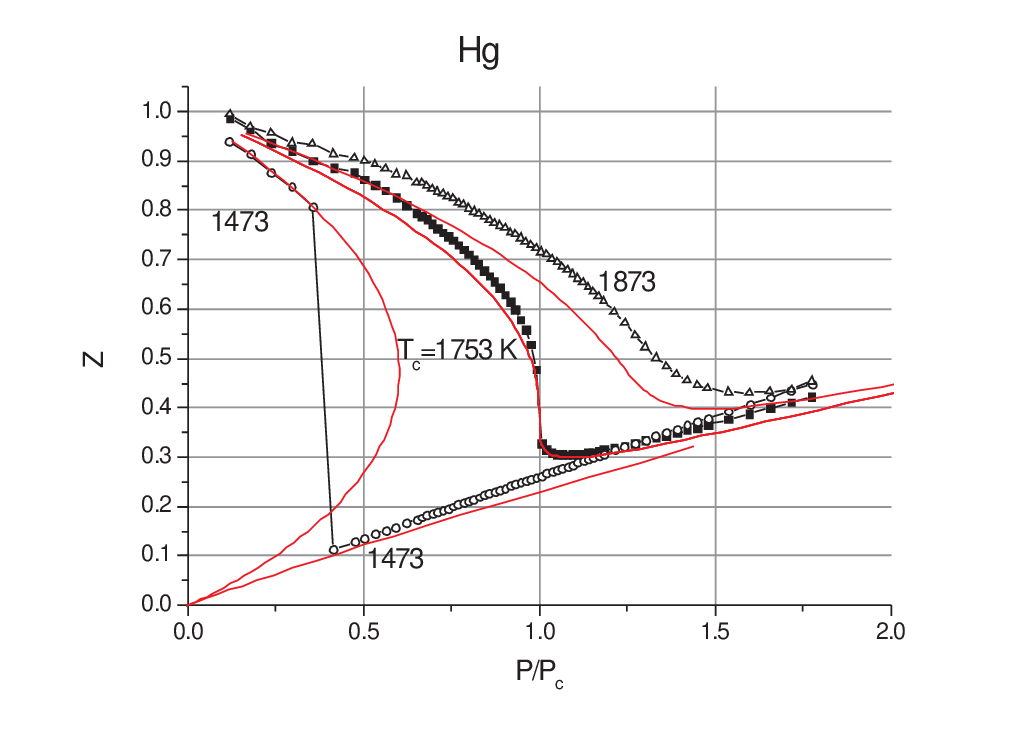}
\end{center}
\caption{The Hougen--Watson diagram for mercury. Experimental data (from the sources
W.~Gotzlaff, G.~Schonherr, F.~Hensel, Z. Phys. Chem. Neue Fol. {\bf156} 219
(1988) and  W.~Gotzlaff., Ph.~D. Thesis, University of Marburg, 1988) are
equipped with symbols. The thick lines correspond to the van der Waals equation
for the related temperatures.}
\end{figure}

\newpage
\clearpage

The \ author \ is \ grateful \ to \ V.\,V.~Brazhkin, A.\,E.~Gekhman, B.\,V.~Egorov, D.\,Yu.~Ivanov,
Yu.\,M.~Kagan, A.\,G.~Kulikovskii, G.\,A.~Martynov, I.\,V.~Melekhov, D.\,S.~Minenkov, V.\,N.~Ryzhov,
A.\,S.~Kholevo, and A.\,V.~Chaplik for fruitful discussions. The author is especially indebted to
Professor~V.\,S.~Vorob'ev who had verified all graphs and and carried out an entire series of the
most important comparisons of theoretical results with experimental data.

\end{document}